\documentclass[fleqn,usenatbib]{mnras}
\usepackage{newtxtext,newtxmath}
\usepackage[T1]{fontenc}
\usepackage{float}

\DeclareRobustCommand{\VAN}[3]{#2}
\let\VANthebibliography\thebibliography
\def\thebibliography{\DeclareRobustCommand{\VAN}[3]{##3}\VANthebibliography}

\usepackage{graphicx}	% Including figure files
\usepackage{amsmath}	% Advanced maths commands
\usepackage{amssymb}	% Extra maths symbols
\usepackage{threeparttable}
\usepackage{xcolor}

\graphicspath{{figures/}}

\title[Plasma scattering screens towards 2005+403]{Probing plasma scattering screens towards the quasar 2005+403 with long-term RATAN-600 observations}

\author[Koryukova et al.]{\parbox{\textwidth}{
T.~A. Koryukova,$^{1,2}$\thanks{E-mail: tatyana.koryukova@gmail.com}
S.~A. Trushkin,$^{3}$
I.~N. Pashchenko,$^{1}$
A.~B. Pushkarev $^{2,1}$
}
\vspace{0.4cm}\\
\parbox{\textwidth}{
$^1$Lebedev Physical Institute of the Russian Academy of Sciences, Leninsky prospekt 53, 119991, Moscow, Russia\\
$^2$Crimean Astrophysical Observatory, Nauchny 298409, Crimea, Russia\\
$^3$The Special Astrophysical Observatory of the Russian Academy of Sciences, Nizhnij Arkhyz, 369167, Russia\\}}

\date{Accepted 2025 ? ?. Received 2025 ? ?; in original form 2025 March 28}

\pubyear{2025}

\begin{document}
\label{firstpage}
\pagerange{\pageref{firstpage}--\pageref{lastpage}}
\maketitle

% Abstract of the paper
\begin{abstract}

We continue investigating the observed properties of the quasar 2005$+$403 seen through the highly turbulent plasma in the Cygnus region. Our earlier study \citep{Koryukova2023} revealed a great influence of propagation effects on the observed data, e.g. numerous episodes of multiple imaging formation, angular broadening of the source size, and the detection of an extreme scattering event (ESE) in 2019, making it a good probe for thermal plasma in the interstellar medium (ISM). We report the first detection of multi-frequency and multi-epoch ESEs revealed with the RATAN-600 daily light curves of the quasar 2005$+$403. The most prominent ESE flux density modulations are found in 2011, 2015, and 2020 and each lasted about 4--5~months. 
We fitted the detected ESEs jointly at 4.7/4.8, 7.7/8.2, and 11.2~GHz using the models, which allowed us to constrain the angular and linear size of the scattering lens averaged over the three ESEs $0.3\pm0.1$~mas and $0.6\pm0.1$~au, proper motion $8.3\pm0.7$~mas~yr$^{-1}$ and transverse velocity $70.1\pm5.7$~km~s$^{-1}$ assuming that the lens distance is 1.8~kpc, maximum free-electron density on the line of sight $1183\pm124$~cm$^{-3}$ and lens mass $(0.8\pm0.4)\times10^{-15}\,\rm M_\odot$. Using the fitting results, we reconstructed the intrinsic unscattered angular size of the quasar at the ESE epochs. We also report on the first detection of up to six ESEs in a row that occurred in the period of 2015--2016, apparently created by multiple lenses successively crossing the line of sight.

\end{abstract}


%\newpage
%\input{text_of_article}

\section{Introduction}
\label{s:intro}

The interstellar medium (ISM) is a dynamic and diverse environment and is the nexus of numerous phenomena that regulate the formation of stars, the growth of supermassive black holes, and the evolution of galaxies. The ISM is fundamentally inhomogeneous as a result of thermal instabilities \citep{Field1969} and supernova (SN) heating \citep{McKee1977}. Recent magnetohydrodynamic simulations of the turbulent ISM have revealed the ubiquity of thin filament-like structures on small scales that intermittently permeate the diffuse medium \citep{Dong2022, Fielding2023}. It has been shown that ionized shock fronts viewed edge-on create strong refraction effects at radio frequencies \citep{Clegg1988}. 

In our previous paper \citep{Koryukova2023} we observed highly unusual changes in the brightness distribution of the quasar 2005$+$403 at low frequencies (lower than 8.3~GHz). The observed peculiar structural changes were induced by refractive-dominated scattering in the ISM, which creates multiple images of the quasar stretched along the line of constant Galactic latitude. In addition to this phenomenon, we showed the presence of caustic surfaces on the 15~GHz Owens Valley Radio Observatory (OVRO) light curve, indicating that the source was undergoing an ESE caused by the refraction of radio waves passing through the localized interstellar plasma lens \citep{Fiedler1987, Romani1987, Fiedler1994, Clegg_1998}. The line of sight towards the quasar 2005$+$403 passes through the Cygnus region, which is known to host highly turbulent interstellar medium (e.g. \citealt{Mutel1988, Fey1989, Mutel1990, Gabani2006}). The Cygnus region itself is characterized by extensive bright radio emission and encompasses star formation regions \citep{Sridharan2002, Beuther2002, Motte2007}, ultra-compact H\,II regions \citep{Downes1966, Wendker1991, Cyganowski2003}, OB stars \citep{Wright2010}, and supernova remnants \citep{Uyaniker2001}.

ESEs that were first discovered in quasars about half a century ago \citep{Fiedler1987} have presented a long-standing mystery in observations of radio sources. The first reported ESEs \citep{Fiedler1987a} identified on the light curves of the quasars 0954$+$658, 1502$+$106, and 1611$+$343. These events were qualitatively similar in shape, each having a flat-bottomed minimum with surrounding maxima (see fig.~1 in \citealt{Fiedler1987}). Only one of these events (of the quasar 0954$+$658) was apparent at the higher frequency observed (8.1~GHz), where the light curve shows four very strong spikes during the flat-bottomed minimum at 2.7~GHz. The nature of the scattering screens that cause ESE remains unknown. However, e.g. \cite{Walker2007} discussed that these events may be caused by ionised gas associated with selfgravitating, au-sized gas clouds moving at hundreds of km~s$^{-1}$ relative to the diffuse ISM and causing strong shocks.

Light curves of compact extragalactic radio sources are characterized by variability on a wide range of timescales from hours (e.g. \citealt{Witzel1986, Heeschen1987, Marchili2011}) to years (e.g. \citealt{Ulrich1997}). The observed variability can be caused by both intrinsic effects related to the emergence and propagation of new VLBI components or effects of external influence such as scattering in the ISM associated with, e.g. the passage of a plasma lens through the line of sight. It has recently been shown that the Sun can directly influence the observed light curves of radio sources on time scales of days \citep{Marchili2011} to months and years \citep{Marchili2024}. Thus, \cite{Marchili2024} showed that all the systematic variability of AGNs in their analysis demonstrated a strict correlation to solar elongation, as the flux density minima occurred either at minimum or at maximum solar elongation. Also, scenarios of local propagation effects (i.e. interplanetary scintillation) or instrumental effects were discussed in that study. Notably, Sun-related modulations on the light curves of background radio sources have the same signature as ESE. In this paper, we test the scenario of Sun-related variability (SRV) on the detected ESE candidates.

Throughout the paper, by the term 'core' we traditionally mean the apparent origin of AGN jet, which is typically the brightest and most compact feature in VLBI images of blazars \citep[e.g.][]{Lobanov98} if not strongly affected by scattering. All position angles are given in degrees from north through east. The spectral index $\alpha$ is defined as $S_\nu\propto\nu^\alpha$, where $S_\nu$ is the flux density measured at observing frequency $\nu$.

\section{Data in use}
\label{s:data}
The extragalactic radio source 2005$+$403 (J2007$+$4029) is a low-spectral peaked quasar at redshift $z=1.736$ \citep{Boksenberg76}. The analysis of parsec-scale jet kinematics shows superluminal motion with maximum apparent speed $9.76c$ \citep{Lister19}. The source Doppler-factor inferred from brightness temperature analysis is quite low, $\delta=3.1$ \citep{Homan21}, which is likely the reason that it has not been detected at high energies \citep{Lister15}. The quasar is at low Galactic latitude $b=4\fdg3$ towards the Cygnus region ($l=76\fdg8$).

\subsection{Multi-frequency RATAN-600 monitoring at 4.7/4.8, 7.7/8.2, and 11.2~GHz.}
\label{sec:mfratandata}

We used the results of long-time monitoring of the microquasars (Trushkin et al., in prep.), when in addition to these targets, two quasars 2005$+$403 and 2013$+$370 have been observed as calibrators with the RATAN-600 radio telescope. Almost daily observations of 2005$+$403 were carried out at central frequencies 1.25, 2.3, 4.7, 8.2, 11.2, 21.7, and 30~GHz during 2010--2024. The radiometers are equipped with modern low-noise HEMT amplifiers. The flux density of 2005+403 changed from 1.5 to 5~Jy in 2010--2024. Typical values of the relative errors of the flux densities are at a level of 3--5 per~cent at frequencies 2.3--11.2~GHz, sometimes increasing from various interference. The central frequencies of the radiometers changed from 7.7 to 8.2~GHz in 2013 and 4.8 to 4.7~GHz in 2019 after modernization of the equipment in order to leave the local interference zone.

The common accepted flux scales \citep{Baars1977,Ott1994}, included secondary calibrator NGC7027 (J2107$+$42) were used for accurate flux density measurements of microquasars and quasars. As a result, the multi-frequency light curves of 2005$+$403 have small relative errors (see \autoref{fig:ratan_lcs}, upper panel).

The low-frequency (1.25 and 2.3~GHz) measurements being contaminated with interference were excluded from further analysis. Averaging the flux densities ($S$) over a range of $\pm0.1$~yr for each measurement, we filtered out data points with deviations from this averaged light curve exceeding $2\sigma_{\rm S}$, where $\sigma_{\rm S}$ is the uncertainty of the flux density measurement. At each frequency, the number of measurements was reduced by no more than 3~per~cent.

\subsection{VLBA data}

We used the open MOJAVE program\footnote{\url{https://www.cv.nrao.edu/MOJAVE}} data of \href{https://www.cv.nrao.edu/MOJAVE/sourcepages/2005+403.shtml}{2005+403} at 15.4~GHz. Data reduction, including initial amplitude and phase calibration, was performed with the NRAO Astronomical Image Processing System \citep[{\sc aips},][]{Greisen2003} following standard techniques. CLEANing \citep{Hongbom1974}, phase and amplitude self-calibration \citep{Jennison1958, Twiss1960} were performed in the Caltech {\sc difmap} package \citep{Shepherd_1997}. The final maps were produced by applying natural weighting of the visibility sampling function. We restored maps with a circular beam to avoid the influence of the beam shape on the source brightness distribution, which is especially useful in cases with anisotropic scattering of the source. The circular beam is obtained by averaging the major and minor axes of the elliptical restoring beam. The uncertainty of the obtained flux densities is assumed at the level of 5~per~cent (e.g. \citealt{Kovalev2005,Lister2018}). The source structure was model-fitted in the spatial frequency plane $(u,v)$ with {\sc difmap} procedure {\sc modelfit} using a limited number of circular Gaussian components that, after being convolved with the restoring beam, adequately reproduce the constructed brightness distribution.

\section{Methods and models in use}
\label{s:methods_and_models}

\subsection{Band-shaped 1D plasma lens model}
\label{s:Fiedler_model}

\cite{Fiedler1987} first discovered atypical flux density variations on the light curves of quasars and attributed them to scattering effects in the ISM. They defined the unusual variations as 'extreme scattering events', reflecting a statistical approach in modeling these events, in which ray path distortions are produced by a large number of electron density fluctuations confined to a localized region of the ISM. ESEs identified at 2.7~GHz did not have counterparts at higher frequencies. Only in one case, for the source 0954$+$658, ESE appeared in 8.1~GHz light curve too. Variations at two frequencies were very dissimilar, reflecting the probing of different scales of turbulent medium. The scattering screen (lens) contains electron density fluctuations that cause an angular broadening of the ray bundles arriving from a background radio source. The angle $\theta_{\rm b}$ through which the rays are dispersed scales with frequency as $ \theta_{\rm b} \propto \nu^{-2}$.

The model developed by \cite{Fiedler1994} assumes that the lens can be approximated by a thin 1D screen, where the broadening angle, the size of the background source, and the screen itself are all of comparable angular extent \citep{Fiedler1994}. The basic function of the screen is to broaden the angular distribution of rays from a background source. In their model the caustic structures are de-emphasized due to angular broadening of the ray paths, but enhancements in the background source flux density are produced by edge effects. They adopted a band-shaped lens as projected on the sky. This geometry is consistent with an edge-on view through a nearly planar lens. The source brightness distribution is assumed to be Gaussian with a full width at half-maximum (FWHM) of $\theta_{\rm s}$. For a point source, it is assumed that electron density inhomogeneities throughout the lens (with angular size $\theta_{\rm l}$) redistribute radiation randomly over a range of angles characterized by the angular width $\theta_{\rm b}$ which is constant across the lens. Flux redistribution resulting from an ensemble of refractive material within the lens is also assumed to follow a Gaussian distribution. The resulting light curve is derived by integrating the redistributed radiation for all possible incidence angles owing to a background source of finite angular size and combining this with the direct flux. Thus, the resulting flux density $S(x)$ is the sum of three integrals in the lens plane:

\begin{gather}
    S(x) = N[c(x - \theta_{\rm l,s})/\sigma_{\rm s},\,c(x + \theta_{\rm l,s})/\sigma_{\rm s}]
    +N[c(\theta_{\rm l,s} - x),\,+\infty]+ \nonumber\\
    +N[-\infty,\,c(- \theta_{\rm l} - x)], \nonumber\\
    \sigma_{\rm s} = \sqrt{1+\theta_{\rm b,s}^2}, \\
    N[a,b] = \frac{1}{\sqrt{2\pi}} \int_{a}^{b} e^{-x^2/2}\,dx, \nonumber 
    \label{eq:Fiedler_model}
\end{gather}

\noindent
where $\theta_{\rm l,s}$ is the lens angular size in $\theta_{\rm s}$ units, $\theta_{\rm b,s}$ is the broadening angle in $\theta_{\rm s}$ units, $x$ is the coordinate in the lens plane, $c$ is the constant defined as $c = \sqrt{8\ln2}$. We added a second index $s$ to the parameter if it is measured in relative units of the source size. For more details, see Appendix~A in \cite{Fiedler1994}.

\subsection{Gaussian plasma lens model}
\label{s:Clegg_model}

\cite{Clegg_1998} presented the geometrical optics for refraction of a distant background radio source by an interstellar plasma lens, with specific application to a lens with a Gaussian profile of free-electron column density. The refractive properties of the lens are specified completely by a dimensionless parameter $\alpha$, which is a function of wavelength ($\lambda$), amplitude of free-electron column density through a one-dimensional plasma lens ($N_0$), lens-observer distance ($D$), and diameter of lens transverse to the line of sight ($a$):

\begin{gather}
    \alpha = 3.6 \left(\frac{\lambda}{\rm 1\,\,cm}\right)^2\left(\frac{N_0}{\rm 1\,\,cm^{-3}\,\,pc}\right)\left(\frac{D}{\rm 1\,\,kpc}\right)\left(\frac{a}{\rm 1\,\,au}\right)^{-2},
	\label{eq:Clegg_alpha}
\end{gather}

\noindent
where $N_0$ is defined as $N_{\rm e} = N_0 \exp[-(x/a)^2]$, $x$~-- is a coordinate in the lens plane. The Gaussian distribution is a convenient function for describing a lens that is localized to some region of characteristic size $a$. Because of the non-uniform free-electron column density $N_{\rm e}(x)$, the lens refracts the ray through the angle $\theta_{\rm r}$. The total observed flux density of the background source $S(u', \alpha)$, where $u' = x'/a$ is a coordinate in the observer plane in lens size units, is obtained by integrating the product of the gain factor $G_{\rm k}$ and the source brightness distribution $B(\beta_s)$ over all $\beta_{\rm s} = \theta_{\rm s}/\theta_{\rm l}$ (further $\theta_{\rm s,l}$), and summing the result of this integration over the $n$ images of the background source:

\begin{gather}
     S(u', \alpha) = \rm \sum_{k = 1}^{n} \int_{-\infty}^{+\infty} B(\beta_s)G_{\rm k}(u', \alpha, \beta_s)\,d\beta_s.
	\label{eq:Clegg_tot_int}
\end{gather}

\noindent
The gain factor $G_{\rm k}$ implies interference of refracted rays. The Clegg's model allows us to estimate the additional parameters of the lens, i.e. the maximum free-electron density on the line of sight as $n_{\rm e} = N_0/a$, as well as the mass of the lens as $M_l = m_{\rm p} n_{\rm e} a^3$, where $m_{\rm p}$ is the mass of the proton. For more details, see section~2 in \cite{Clegg_1998}.

\subsection{Fitting the model to data}
\label{sec:fitting}

The ESE models operate with relative angular sizes of the screen and source, i.e. $\theta_{\rm l,s}$, $\theta_{\rm b,s}$ in Fiedler's and $\theta_{\rm s,l}$ in Clegg's. To interpret the results of modeling we applied the estimates of our previous work, which allowed to derive the observed size $\theta_{\rm b}$ of 2005$+$403 at any frequency following:

\begin{gather}
    \theta_{\rm b}^2 = \left(\theta_{\rm int}\nu^{-1}\right)^2 + \left(\theta_{\rm scat}\nu^{-k_{\rm scat}}\right)^2 
	\label{eq:observed_size},
\end{gather} 

\noindent
where $\theta_{\rm int} = 4.2\pm1.5$~mas and $\theta_{\rm scat} = 70.1\pm5.7$~mas (intrinsic and scattered sizes at 1~GHz, respectively), $k_{\rm scat}$ is the scattering index estimated as $2.01\pm0.13$. In order to estimate these parameters we used VLBI core sizes of 2005$+$403 measured at frequencies ranging from 0.61 to 43.2~GHz (see sec.~4.3 in \citealt{Koryukova2023} for the details).

We introduced additional parameters to the models, namely: $\mu$ is the proper motion of the lens ($\rm mas\,\,yr^{-1}$) for scaling the time axis of the light curve to coordinates in the observer plane; $\delta$ is the time shift to accommodate the inaccuracy of determining the epoch of minimum of the ESE curve; $f_{\rm core}$ is the core dominance or fraction of the flux density that was scattered. We consider the observed flux density as the sum of lensed and unlensed parts, i.e. $ S_{\rm obs} = S_{\rm scat} + S_{\rm unscat} = S_{\rm unlensed} \times f_{\rm core}+S_{\rm unlensed}\times(1 - f_{\rm core})$, where $S_{\rm unlensed}$ was determined as the median flux density level for the event. Additionally, each light curve in the model had its own slope parameter ($m_{\rm i}$). The slope on each light curve on scale of 4--5 months is well approximated by a linear trend. On longer time scales, it would be more correct to use the approach described by \cite{Valtaoja1999} or apply a boxcar averaging technique within $\pm0.1$~yr to correct the long-term trend (see sec.~6. in \citealt{Koryukova2023}).

The models of flux modulation described in \autoref{s:Fiedler_model} and \autoref{s:Clegg_model} consist of several main parameters ${\theta_{\rm Fiedler}} = \{\theta_{\rm l,s_i}, \theta_{\rm b,s_i}, \mu_s, f_{\rm{core},i}\}$ and ${\theta_{\rm Clegg}} = \{\theta_{\rm s,l_i}, \mu_l, \alpha_i, f_{\rm{ core},i}\}$. Some of the parameters are degenerate -- $\theta_{\rm l,s_i}$ and $\mu_s$ or $\theta_{\rm b,s_i}$ and $f_{\rm{ core},i}$ for Fiedler's model, $\theta_{\rm s,l_i}$ and $\mu_l$ or $\alpha_i$ and $\mu_l$ for Clegg's. The degeneracy of the model parameters leads to a large uncertainty in the obtained estimates of the lens and source parameters. By setting priors on the parameters, the influence of degeneracy can be reduced. We used uniform priors, the bounds on which are given in \autoref{tab:model_parameters_fiedler}. Thus, to infer the model parameters from the observed multi-frequency light curves, it is essential to employ the Bayesian approach \citep{bayes,laplace}, see also \citep{2008ConPh..49...71T} for a more recent review. In Bayesian statistics, the prior probability distribution of the model parameters $\pi(\bold{\theta})$ is updated by the observed data $\bold{y}$ through the likelihood $L(\bold{y} | \bold{\theta})$ to obtain the posterior distribution $P(\bold{\theta} | \bold{y})$:

\begin{equation}
    P(\bold{\theta} | \bold{y}) = \frac{L(\bold{y} | \bold{\theta}) \pi(\bold{\theta})}{\int L(\bold{y} | \bold{\theta}) \pi(\bold{\theta}) d\bold{\theta}}
\label{eq:posterior}
\end{equation}
Here the normalization constant in the denominator is the evidence $Z$ (or marginalized likelihood) - the probability of the model given the data. 

In principle, the samples from the posterior distribution can be directly obtained by e.g. Markov Chain Monte Carlo (MCMC) methods. However, for complex multi-modal distributions the \textit{Nested sampling} \citep{skilling2004nested} can be more effective. Nested Sampling aims to estimate the evidence $Z$ by integrating the prior $\pi$ to nested levels of constant likelihood. This effectively converts multi-dimensional integral to one dimensional integral of a monotonic function - likelihood over the prior mass enclosed within a likelihood contour \citep{10.1214/06-BA127}. The by-product of the nested sampling run is the sample from the posterior distribution.

We employed the Nested Sampling algorithm implemented in \texttt{Dynesty} library \citep{speagle2020dynesty,sergey_koposov_2024_12537467}. Our opted nested sampling implementation in \texttt{Dynesty} starts with a predefined number of the \textit{live} points (parameter \textit{nlive} $= 512$) drawn from the prior and performs sampling from the constrained prior by considering only live points with a likelihood greater than a current threshold value (level). The latter is the lowest likelihood of the current live point set, which monotonically increases during the run by removing the corresponding live point. To replenish the ``dead'' points, new live points are generated from the evolving system of the bounding ellipsoids \citep{2009MNRAS.398.1601F} which enclose the current set of live points (default parameter \textit{bound} = `multi'). A more detailed description of the nested sampling implementation could be found in the corresponding \texttt{Dynesty} release papers.

The scattering lens models used in this paper are sensitive to data variations. In order to avoid the influence of outlier data points on the fitting result, we employ the robust Student-t likelihood with number of degrees of freedom equal to 2. The Student-t distribution \citep{lange1989robust} is similar to the normal distribution, but has a higher chance of extreme values than the corresponding normal distributions with the same scale parameter.
The estimated flux density uncertainties are significantly larger than the systematical variations in the light curves, suggesting that they are the upper limits. Thus, we introduced additional parameters during the fit, the uncertainty at each frequency band, and marginalized them out when presenting the lens model parameters \citep{2010arXiv1008.4686H}.  

\section{Jet morphology}
\label{s:quasar_properties}

In our previous paper \citep{Koryukova2023} we studied the heavily scattered quasar 2005$+$403 and found:

\begin{enumerate}
    \item numerous episodes of multiple imaging \citep{Clegg_1998}, when 1--3 sub-images of the source were detected;
    \item angular broadening effect \citep{Duffett_Smith1976} that manifested itself at each observing epoch at frequencies ranging from 1.4 to 8.3~GHz;
    \item it is appeared to have undergone an ESE \citep{Fiedler1987} started around 2019 March and lasted for approximately 1.6 months.
\end{enumerate}

\noindent The detection of both diffractive and refractive scattering effects indicates strong turbulence and heterogeneity of the ISM along the line of sight toward 2005$+$403 making it a good probe for studying the properties of intermediate plasma.

\begin{figure}
    \centering
    \includegraphics[width=1\linewidth]{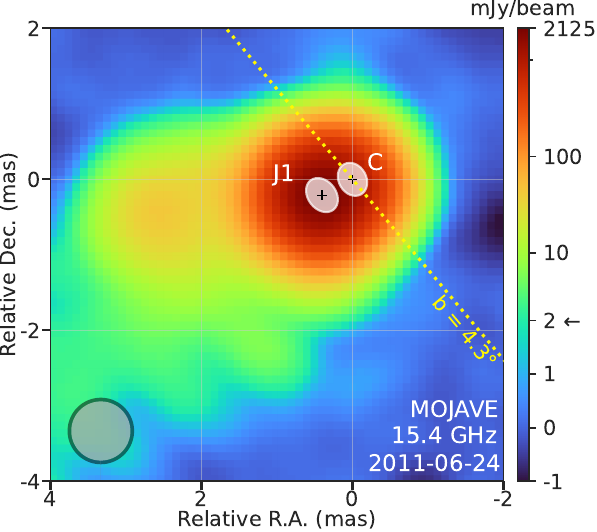}\vspace*{0cm}
    \caption{VLBA MOJAVE total intensity map of 2005$+$403 at 15.4~GHz map. White ellipses represent the fitted Gaussian components of the core (C) and the innermost jet (J1). We show only two brightest components of the jet that were subject to scattering. The full model consists of five components and presented in \autoref{tab:mojave_model}. The map and model were shifted such that the fitted core position is aligned with the phase center. The colour bar scale transits from linear to logarithmic at 2~mJy~beam$^{-1}$. The gray circle in the lower left corner is FWHM of the corresponding restoring beam. Yellow dotted line is the line of constant Galactic latitude ($b = 4\fdg3$) at $\mathrm{PA}=40\fdg6$. } 
    \label{fig:mojave_maps}

\end{figure}

It has been shown that at relatively high observing frequencies, 23.8 and 43.2~GHz, the source brightness distribution shows a typical one-sided core-jet AGN morphology (see the upper left plots in fig.~2 and fig.~A1 in \citealt{Koryukova2023}). The VLBI-core component located at the apparent jet origin is the brightest and most compact feature. The other jet components progressively weaken with core separation and propagate at $\mathrm{PA}=110^\circ$. At 15.4~GHz the source also shows one-sided core-jet AGN morphology (\autoref{fig:mojave_maps}), but the innermost jet components follow a curved trajectory and their motion is not ballistic, suggesting a spatially curved (helical) jet, but beyond 2~mas core-separation the components seem to move on ballistic trajectories \citep{Gabani2006}. In \autoref{tab:mojave_model}, we present the model fitting results of the data measured at 15.4~GHz in the MOJAVE project and depicted in \autoref{fig:mojave_maps}. The model consists of five Gaussian components including circular and elliptical ones. The errors of the obtained model fit parameters were estimated from the image plane using the analytical approximations from \citealt{Schinzel2012} (see sec.~4.1 in \citealt{Koryukova2023} for details of the method). The model contains components describing the extended jet emission, as well as two bright and compact components at the jet origin (C and J1 in \autoref{fig:mojave_maps}), which apparently show evidence of anisotropic scattering (discussed in \autoref{s:multiple_ese}).

\begin{table*}
	\caption{The results of model fitting using elliptical and circular Gaussian components for the data illustrated in \autoref{fig:mojave_maps}.}
	\centering
	%\begin{tabular}{|*{6}{c|}}
        \begin{tabular}{lcccccc}
    
		\hline
		Component     &  $S$           &  $r$           & $\varphi$         & $\theta$         & Axial ratio & $\Phi$\\
                          & (Jy)           & (mas)          &  ($\degr$)     &  (mas)           &             & ($\degr$) \\
            (1)           & (2)            & (3)            &  (4)           & (5)              &  (6)        & (7) \\
		\hline
            core (C)      & $0.81\pm0.05$  & $0.00$         & \ldots         & $0.45\pm0.05$    & 0.79        &  29.3 \\ 
            jet (J1)      & $2.26\pm0.09$  & $0.45\pm0.02$  & $117.3\pm2.1$  & $0.49\pm0.03$    & 0.76        &  37.7 \\ 
            jet           & $0.25\pm0.04$  & $0.88\pm0.11$  & $117.2\pm7.1$  & $1.08\pm0.22$    & 1.00        &  $0.00$  \\ 
            jet           & $0.10\pm0.02$  & $2.44\pm0.12$  & $99.5\pm2.9$   & $0.93\pm0.24$    & 1.00        &  $0.00$  \\ 
            jet           & $0.05\pm0.04$  & $3.59\pm1.08$  & $128.1\pm16.7$ & $3.03\pm2.16$    & 1.00        &  $0.00$  \\ 
            \hline
	\end{tabular}
         \begin{tablenotes}
            \item The columns are as follows: (1) type of a component; (2) flux density; (3) radial distance relative to the core; (4) PA with respect to the core; (5) FWHM of the component size; (6) axial ratio of the Gaussian component ('1' for circular component, $<1$ for elliptical component); (7) PA of the major axis of the ellipse.
        \end{tablenotes}
	\label{tab:mojave_model}
\end{table*}

At frequencies lower than 8.3~GHz we still detect the jet emission, but the observed quasar morphology is dominated by scattering effects and can be highly atypical due to anisotropic scattering effects (see the plots fig.~2 and fig.~A1 in \citealt{Koryukova2023}). At low frequencies (1.4 and 2.3~GHz) the jet emission is not visible and the observed brightness distribution of the source is fully formed by scattering and stretched nearly along the line of constant Galactic latitude $\rm b = 4\fdg3$ (at $\mathrm{PA} = 40^\circ$), because of the creation of secondary images of the source (e.g. the lower right plot in fig.~2 in \citealt{Koryukova2023}).

\section{RATAN-600 light curves}
\label{s:lc_analysis}

\subsection{Flux density evolution}

\begin{figure*}
    \centering 
    \includegraphics[width=1\linewidth]{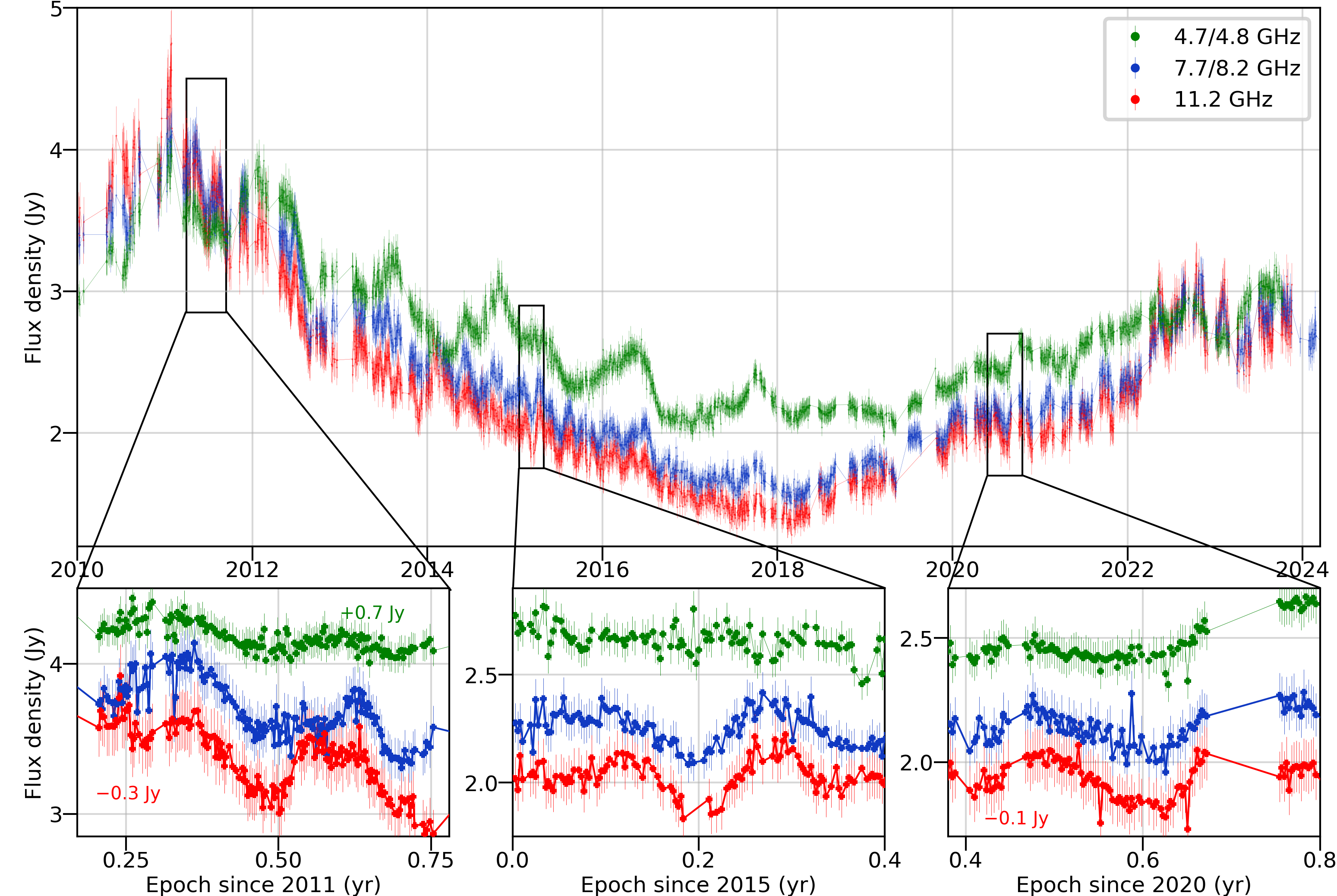}\vspace*{0cm} 
    \caption{Upper panel: The RATAN-600 light curves of 2005$+$403 measured at 4.7/4.8, 7.7/8.2, and 11.2~GHz (green, blue, and red colour, respectively). The relative measurement error is about 3, 4, and 5 per~cent for 4.7/4.8, 7.7/8.2, and 11.2~GHz, respectively. Lower panel: zoom in plots of the ESE candidates. Some light curves shown in the lower panel have been shifted lower or upper to avoid visual interlacing.}
    \label{fig:ratan_lcs}
\end{figure*}

In \autoref{fig:ratan_lcs} we show the RATAN-600 4.7/4.8, 7.7/8.2, and 11.2~GHz light curves. The full RATAN-600 data contain daily observations covering epochs from 2005 December 01 to 2024 April 16. We excluded data before 2010 from the analysis due to significantly lower sampling of observations. The relative measurement error is approximately 3, 4, and 5~per~cent for 4.7/4.8, 7.7/8.2, and 11.2~GHz data, respectively.

\begin{figure}
    \centering
    \includegraphics[width=1\linewidth]{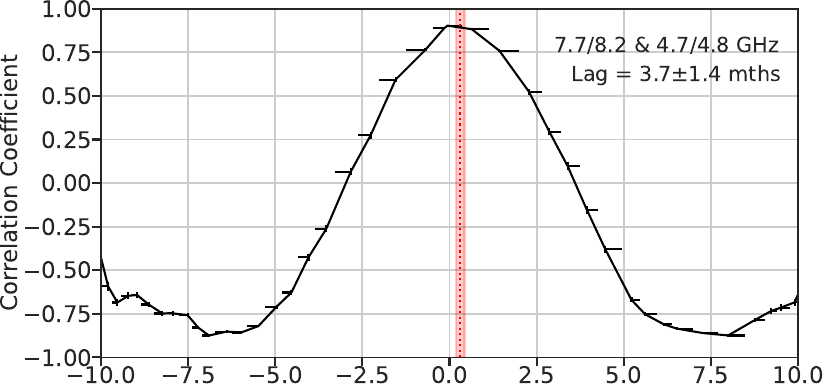}\vspace*{0cm}
    \includegraphics[width=1\linewidth]{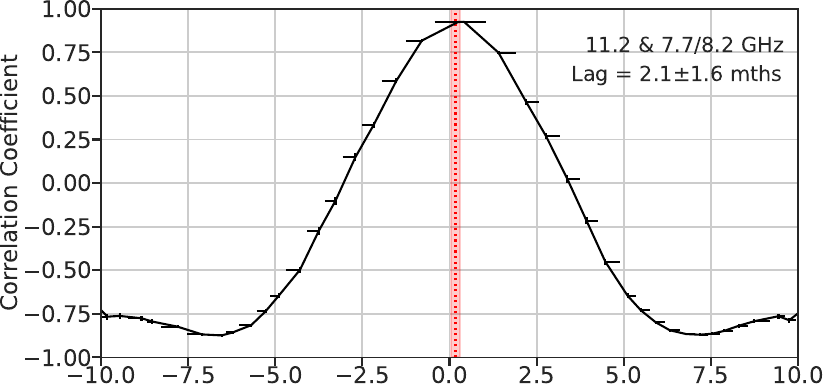}\vspace*{0cm} 
    \includegraphics[width=1\linewidth]{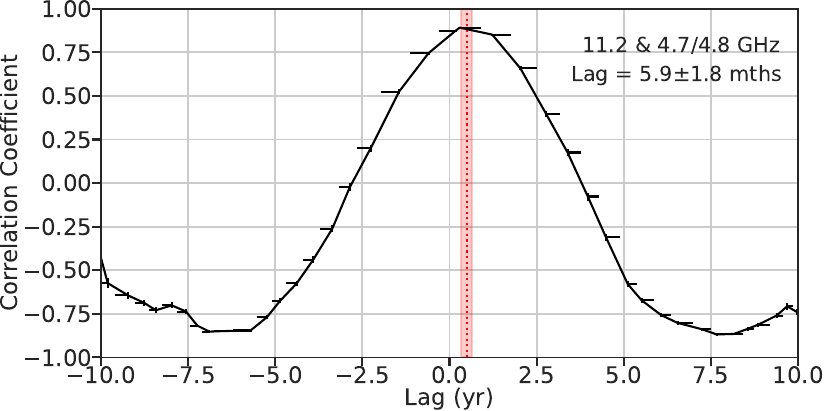}\vspace*{0cm} 
    \caption{zDCF constructed from the RATAN-600 light curves of the quasar 2005$+$403 measured at different frequencies. Upper: emission at 4.7/4.8~GHz lags behind 7.7/8.2~GHz by 3.7 months. Middle: emission at 7.7/8.2~GHz lags behind 11.2~GHz by 2.1 months. Lower: emission at 4.7/4.8~GHz lags behind 11.2~GHz by 5.9 months. The vertical dotted red line depicts the maximum correlation coefficient lag estimated by fitting zDCF output with a normal distribution. The shaded reddish area represents $1\sigma$ confidence interval.}
    \label{fig:zdcf}
\end{figure}

To test the light curves for a time delay between pair of frequencies, we applied the z-transformed discrete correlation function (zDCF; \citealt{Alexander1997}) using Python module called {\sc pyZDCF} developed by \cite{pyzdcf}. It has been developed for robust estimation of cross-correlation function of sparse and unevenly sampled astronomical time series. The zDCF takes into account that the sampling distribution of the correlation coefficient is often far from normal. Thus, a more accurate error estimation other than the standard deviation formula is required. The zDCF binning algorithm shares the idea with adaptive binning (e.g. \citealt{Lott2012}) in the sense that the bin width is varied to make sure that the statistical significance is sufficiently high for each bin. In \autoref{fig:zdcf} we show zDCF constructed for the RATAN-600 light curves of the quasar 2005$+$403.

Fitting the result of zDCF algorithm output with normal distribution we calculated the lag that corresponds to the highest correlation coefficient. The error of the lag was estimated applying bootstrap method. We measured a time delay between 7.7/8.2 and 4.7/4.8~GHz, 7.7/8.2 and 11.2~GHz, as well as 4.7/4.8 and 11.2~GHz. A positive time lag corresponds to leading of the higher frequency data over lower frequency data, which is expected from synchrotron opacity \citep{Pacholczyk1970, BK79}. Generally, the variability of 2005$+$403 shows spectral evolution: a correlation analysis of the light curves using zDCF reveals that variations at 11.2~GHz occur about 2 months earlier than at 7.7/8.2~GHz, and about 6 months earlier than at 4.7/4.8~GHz. The time lag of 3.7 months corresponds to 7.7/8.2~GHz light curve variations leading variations at 4.7/4.8~GHz.

\subsection{Long-term variability}

The RATAN-600 light curves are characterized by variability on different time scales. The longest observed scale is about seven years and is clearly observed at all frequencies. In AGN, the long-term total flux density variations are often related to the emergence and propagation of new bright VLBI components. But the corresponding time scale of a year or a few \citep{Lister2021} is much lower compared to the $\sim13$~yr between two consecutive peaks on 2005$+$403 light curves. 
%(Trushkin, priv. comm.). 
Structural analysis of the 447 brightest radio-loud sources in the northern sky at 15~GHz \citep{Lister2021} showed that variations in the inner jet PA with an average amplitude of $10-50\degr$ on a time scale of about a decade were observed very frequently. Possible explanations for such variations included: (i) orbital motion in a binary black hole system \citep{Begelman1980}; (ii) item precessions of the accretion disk, the rotation axis of which is offset from the rotation axis of the black hole \citep{Lense1918, Thirring1918, Caproni2004}.

\cite{Gabani2006} modeled a one-sided core-jet structure of 2005$+$403 at 15~GHz with five Gaussian components for eight different observing epochs obtained during 1995--2003. The estimated apparent speed of the components range from 6.3 to 16.8~$c$, indicating intrinsic bending of the jet. Their fitted model revealed the bright and highly variable innermost component of the jet at $\sim0.4$~mas core separation. It showed the most pronounced flux density variability (see fig.~4 in \citealt{Gabani2006}). The detected variability in PA and velocity of the jet components, together with the long-term variability evident in the RATAN-600 light curves can potentially be attributed to jet precession \citep{Todorov2023, Kostrichkin2025}.

\begin{figure}
    \centering
    \includegraphics[width=1.01\linewidth]{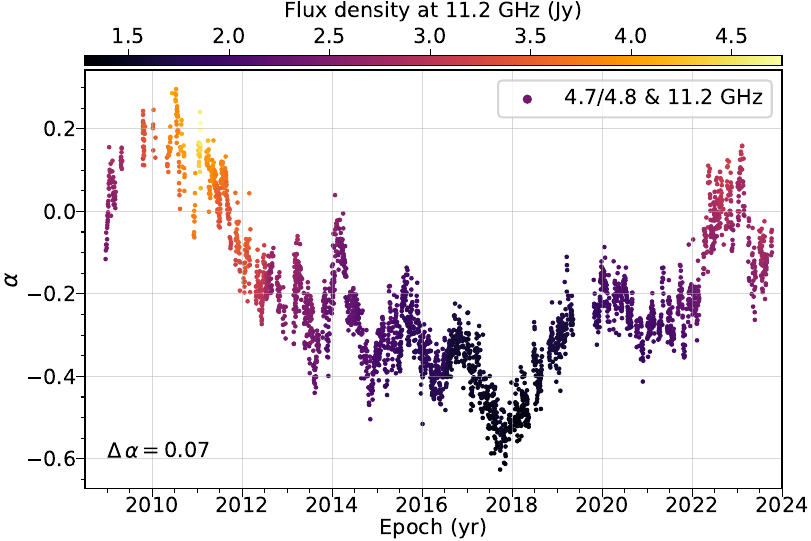}
    \includegraphics[width=1\linewidth]{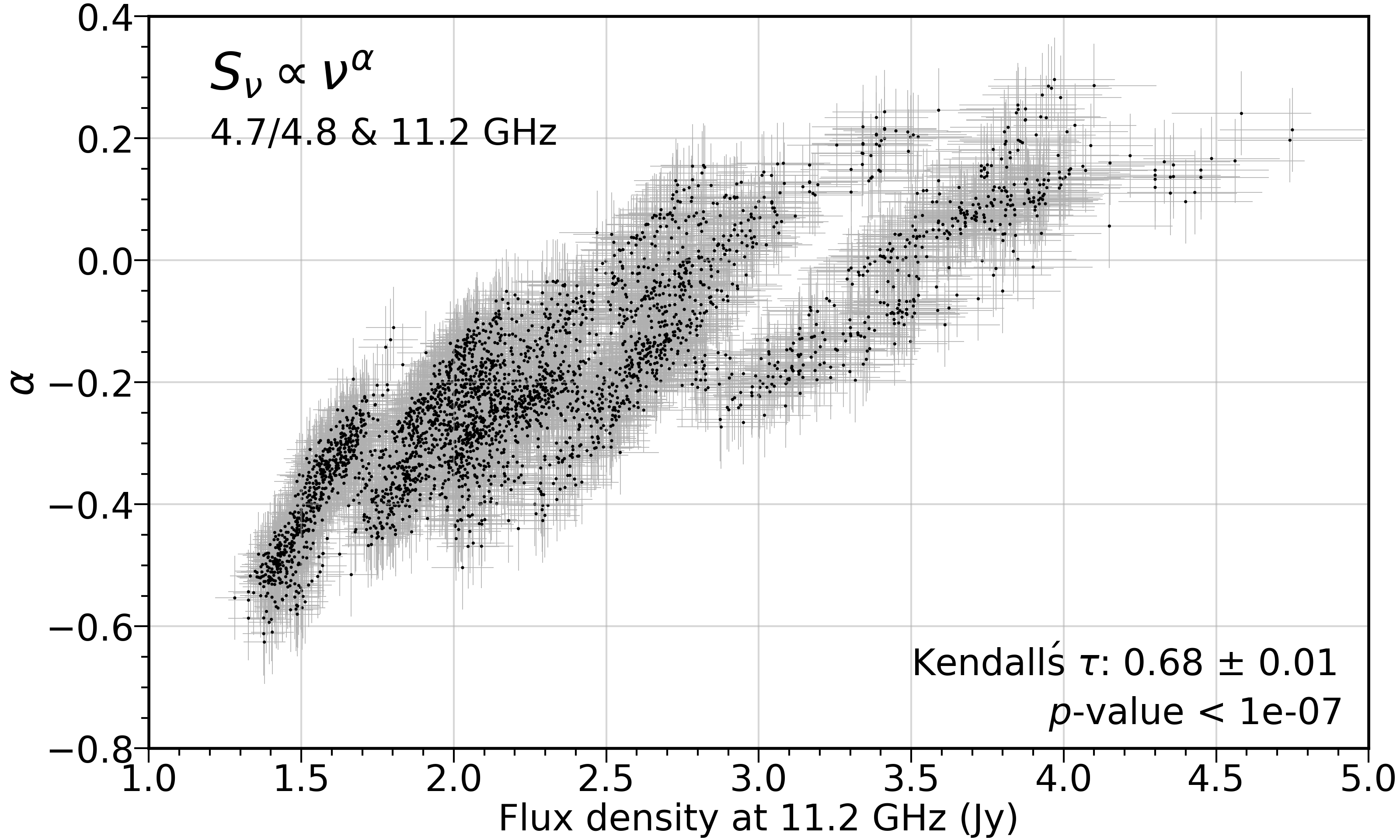}
    \caption{Two-frequency spectral index evolution constructed from 4.7/4.8 and 11.2~GHz data (upper panel).
             Strong correlation between spectral index $\alpha$ measured from 4.7/4.8 and 11.2~GHz data versus flux density at 11.2~GHz (bottom panel).
             Plots based on other frequency pairs are qualitatively similar.}
    \label{fig:spectral_index}
\end{figure}

However, the flux density variability is accompanied by significant spectral variations. In \autoref{fig:spectral_index} (top), we present spectral index $\alpha$ derived from the simultaneous flux density measurements at 4.7/4.8 and 11.2~GHz. The data reveal that the source exhibits a flat spectrum when it is bright, and becomes gradually steeper during the low state. This flatter-when-brighter trend strongly suggests that the long-term variability is primarily driven by intrinsic changes in the source. Specifically, when a flare propagates down the jet, it injects more high-energy emitting electrons leading to a flattening of the radio spectrum. Using Kendall's $\tau_{\rm K}$ non-parametric rank correlation coefficient and randomization test with number of permutations of $10^7$, we found that the spectral index is correlated with the flux density (\autoref{fig:spectral_index}, bottom) at a very high level of significance with $\tau = 0.68\pm0.01$ and $p$-value $<10^{-7}$. It indicates that the light curves trace the source at stages with different energy loss mechanisms \citep{MG87}. This relationship further supports the interpretation that intrinsic processes within the jet are the dominant cause of the observed variability. The errors of $\tau$ was estimated as standard deviations of the distribution formed by bootstrap.

\subsection{Propagation effects on the light curves}

Modulations on the light curves observed over shorter time intervals (1.5~years or less) most likely relate to propagation effects. The wide frequency coverage of our data facilitates a study of the interplay of the frequency-dependent scattering effects, which dominate below 8~GHz. The major contributor to variability of the light curves at gigahertz frequencies is refractive interstellar scintillations (RISS). Variations attributed to RISS appear as low-amplitude fluctuations with decorrelation timescales measured in days-weeks and persist over many years of observation \citep{Rickett1986}. This is generally interpreted, in the case of extragalactic sources, to be the result of the proper motion of small-scale electron density turbulence in the ISM \citep{Cordes1985}. Unusually strong refractive effects in the ISM are responsible for ESE \citep{Fiedler1987, Fiedler1994} and are associated with a passage of localized structures of ionized medium with au-scale transverse dimensions across the line of sight \citep{Cordes86, Clegg1988, Walker2007}. The scattering screen refracts radio waves at different angles forming ray focusing and defocusing regions \citep{Clegg_1998} at a certain distance from the lens plane. These focusing regions are observed as caustic surfaces on the light curves of a background source, and they provide valuable information for studying the physical properties of the intervening scattering screen.

\begin{figure*}
    \centering
    \includegraphics[width=0.45\linewidth]{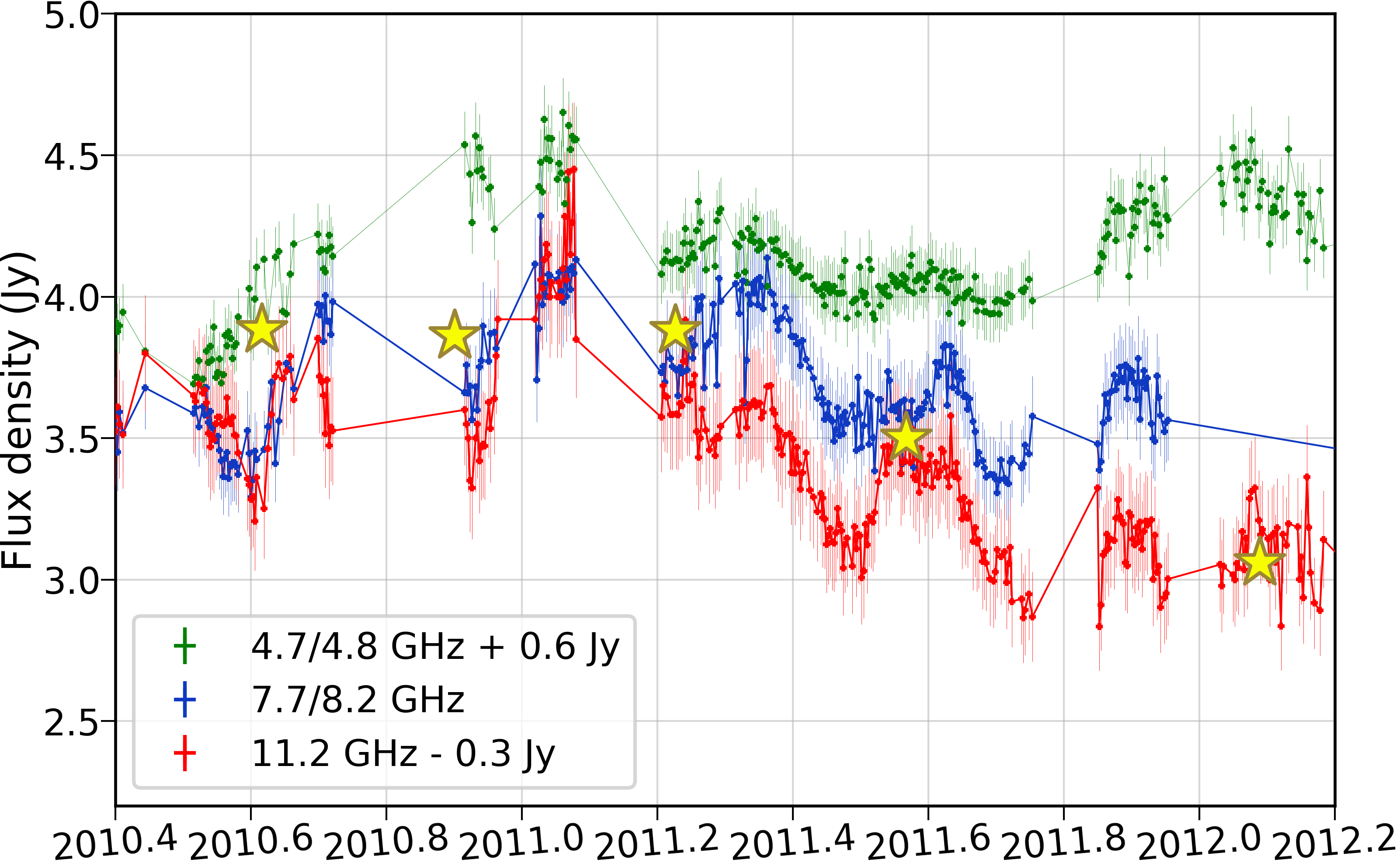}\vspace*{0.1cm}
    \includegraphics[width=0.45\linewidth]{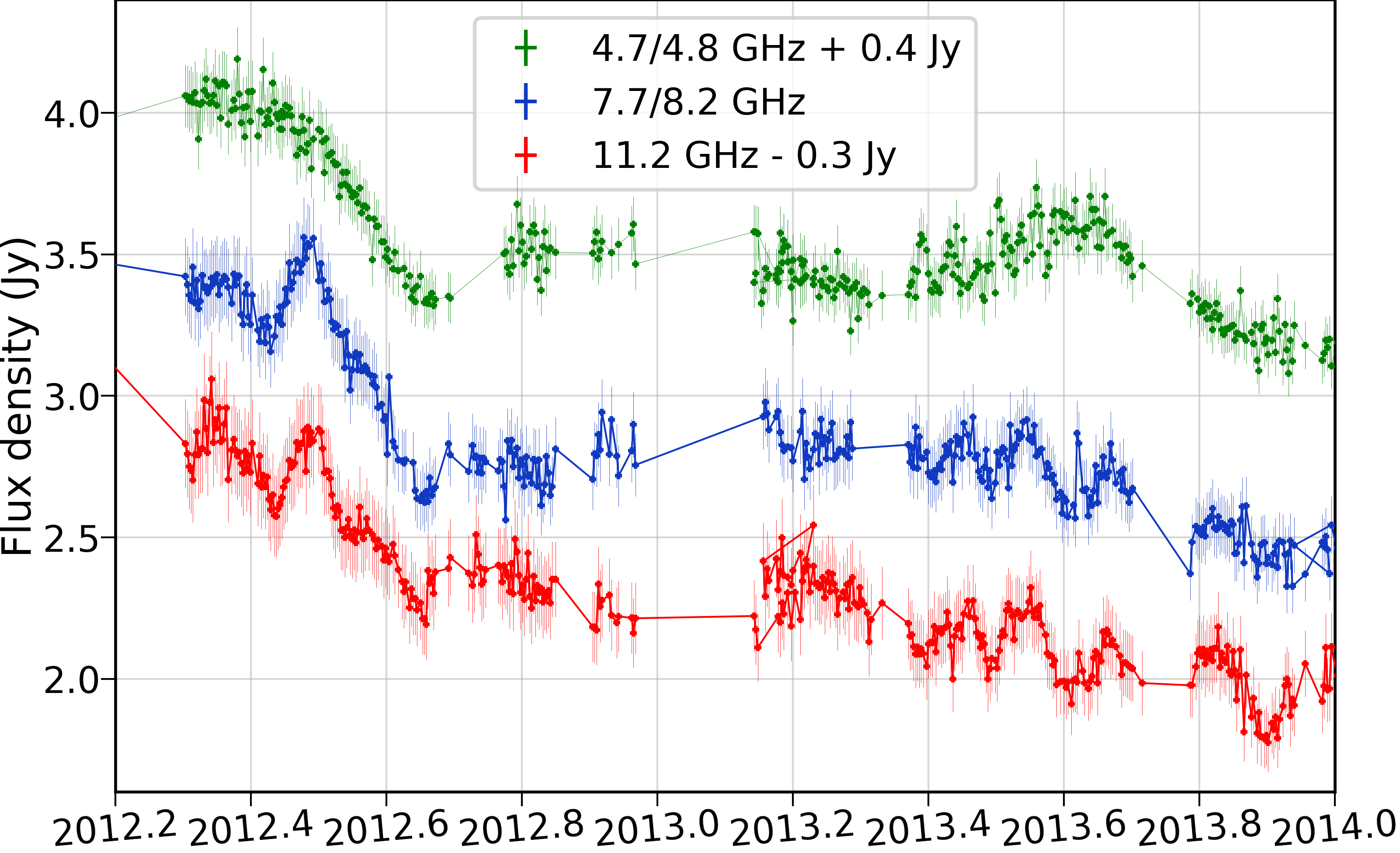}\vspace*{0cm} 
    \includegraphics[width=0.45\linewidth]{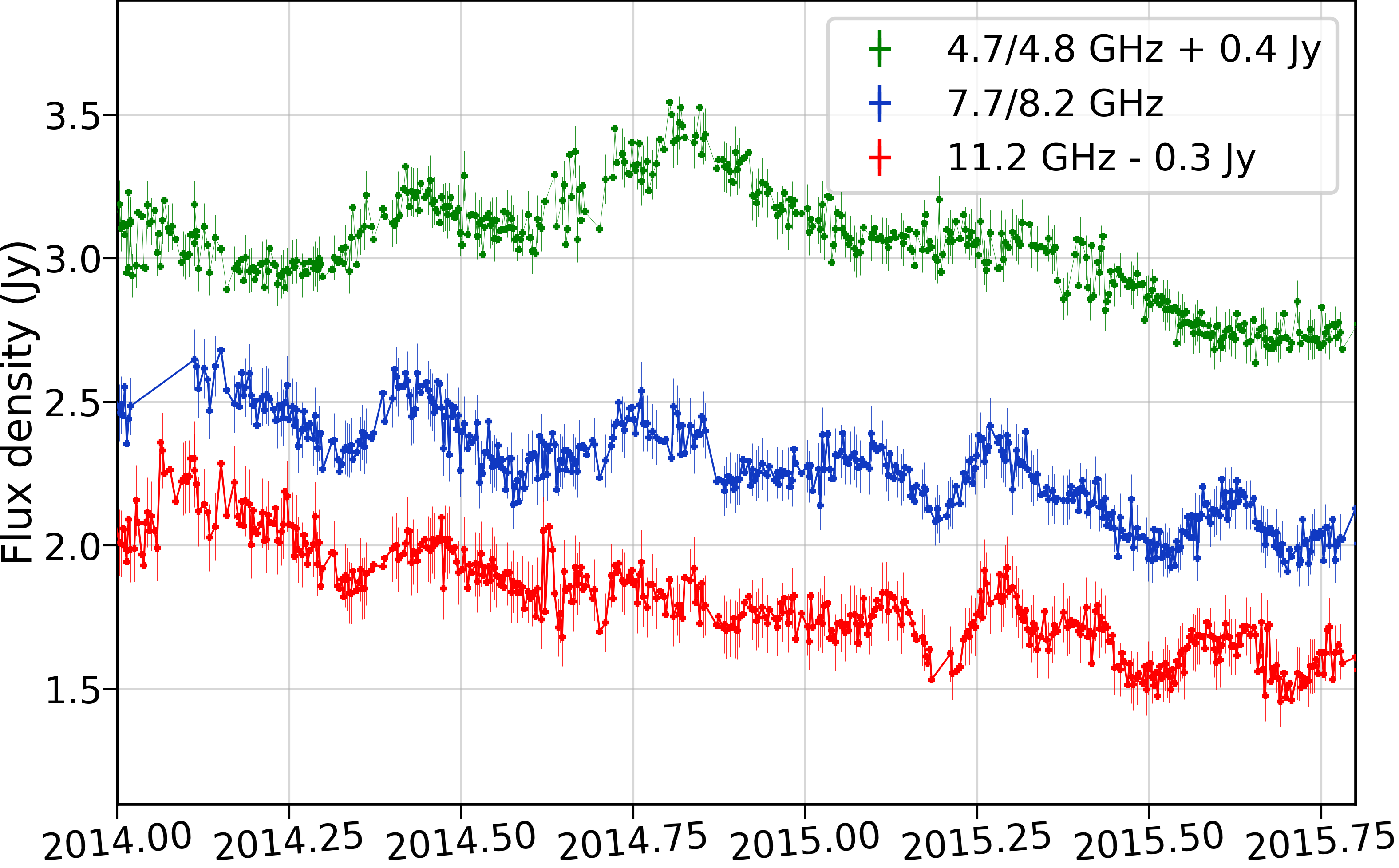}\vspace*{0.1cm} 
    \includegraphics[width=0.45\linewidth]{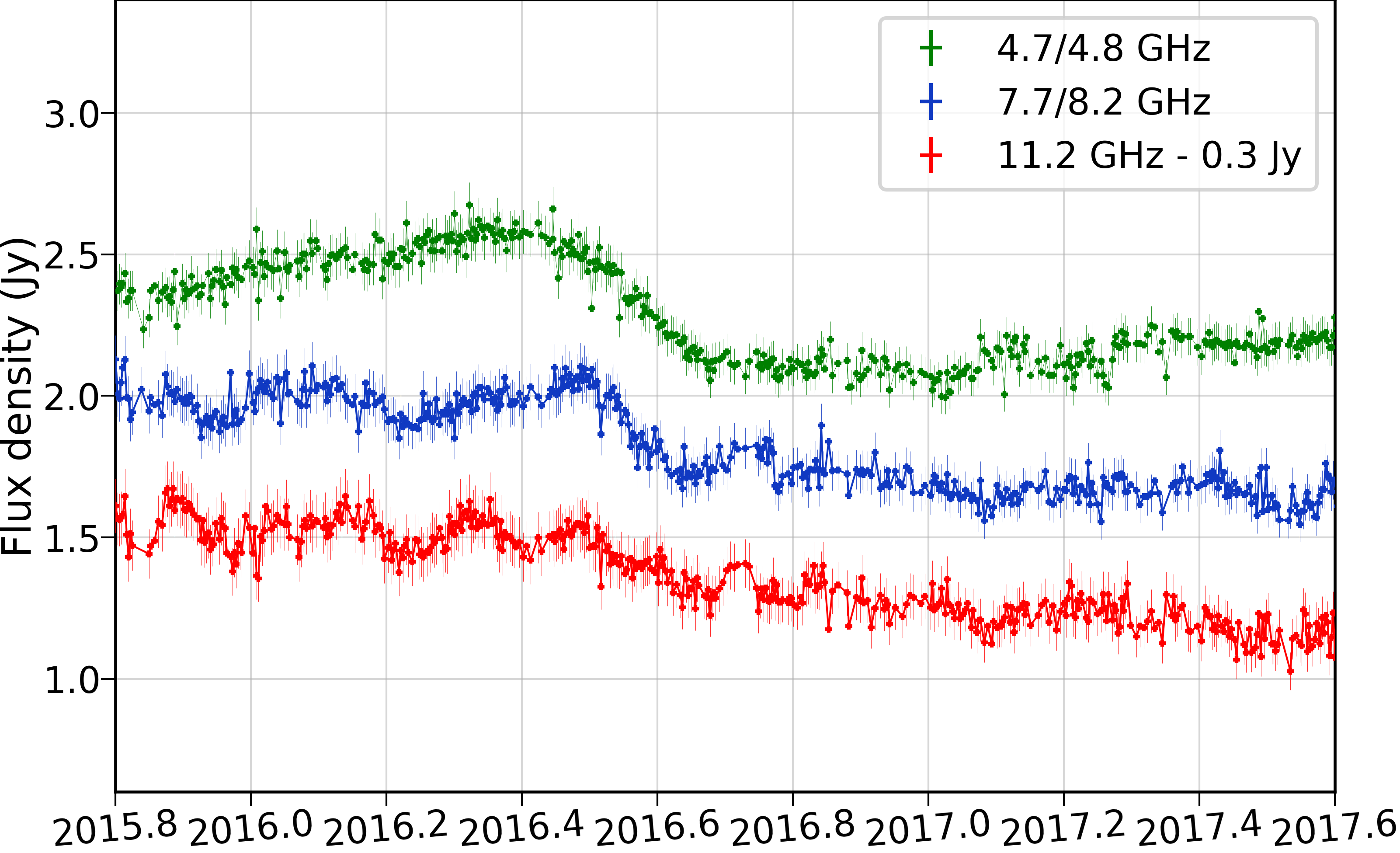}\vspace*{0cm} 
    \includegraphics[width=0.45\linewidth]{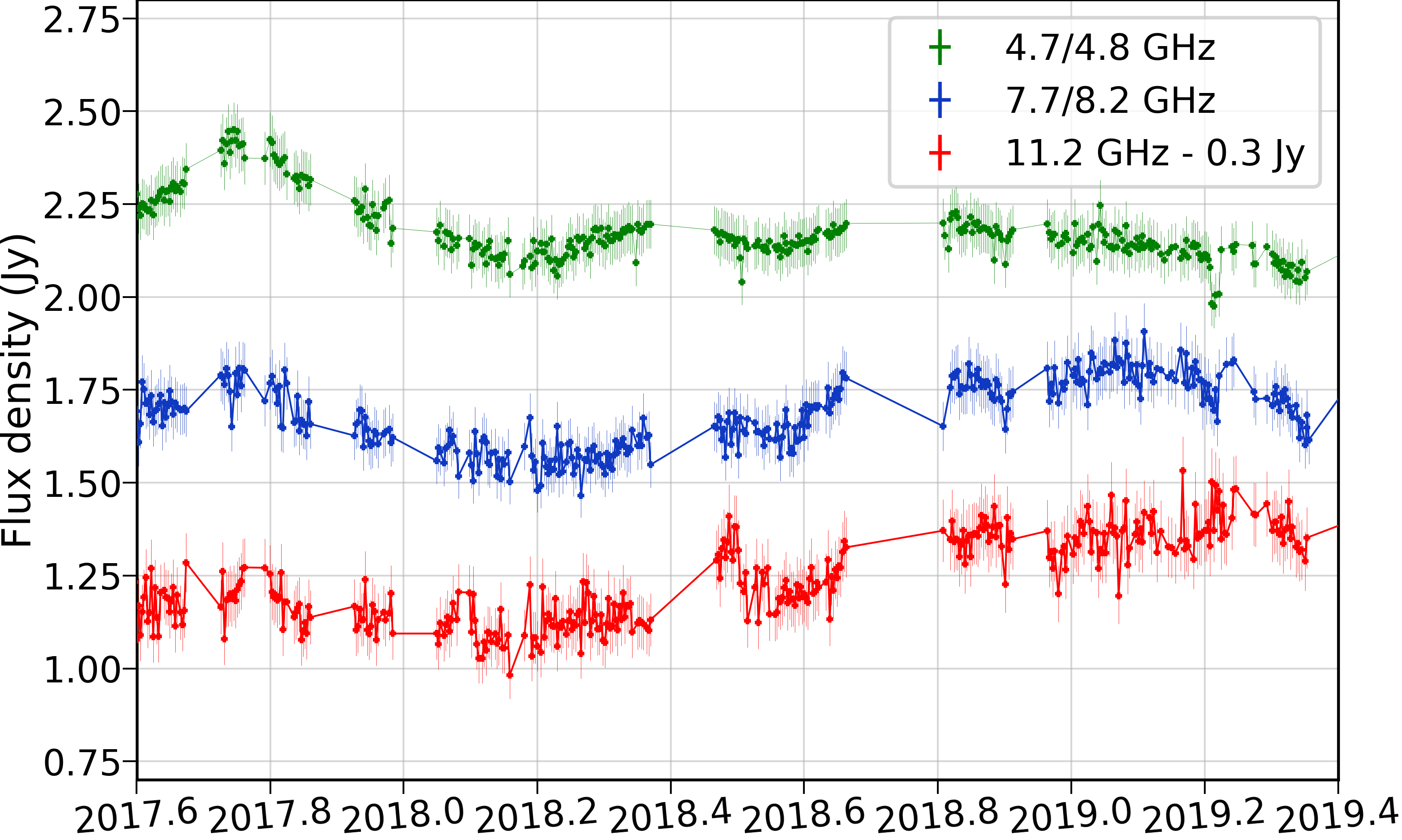}\vspace*{0.1cm}
    \includegraphics[width=0.45\linewidth]{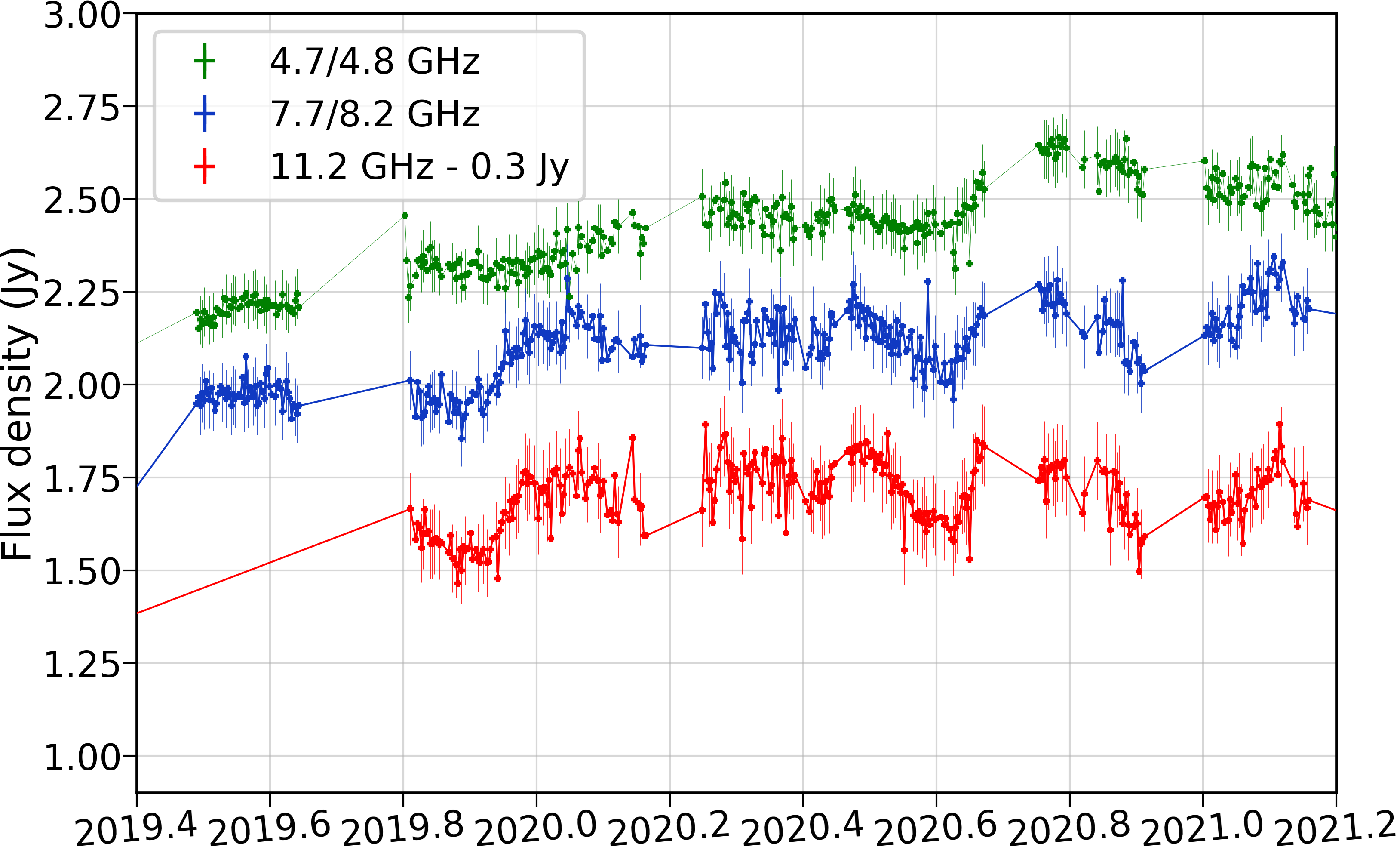}\vspace*{0cm} 
    \includegraphics[width=0.45\linewidth]{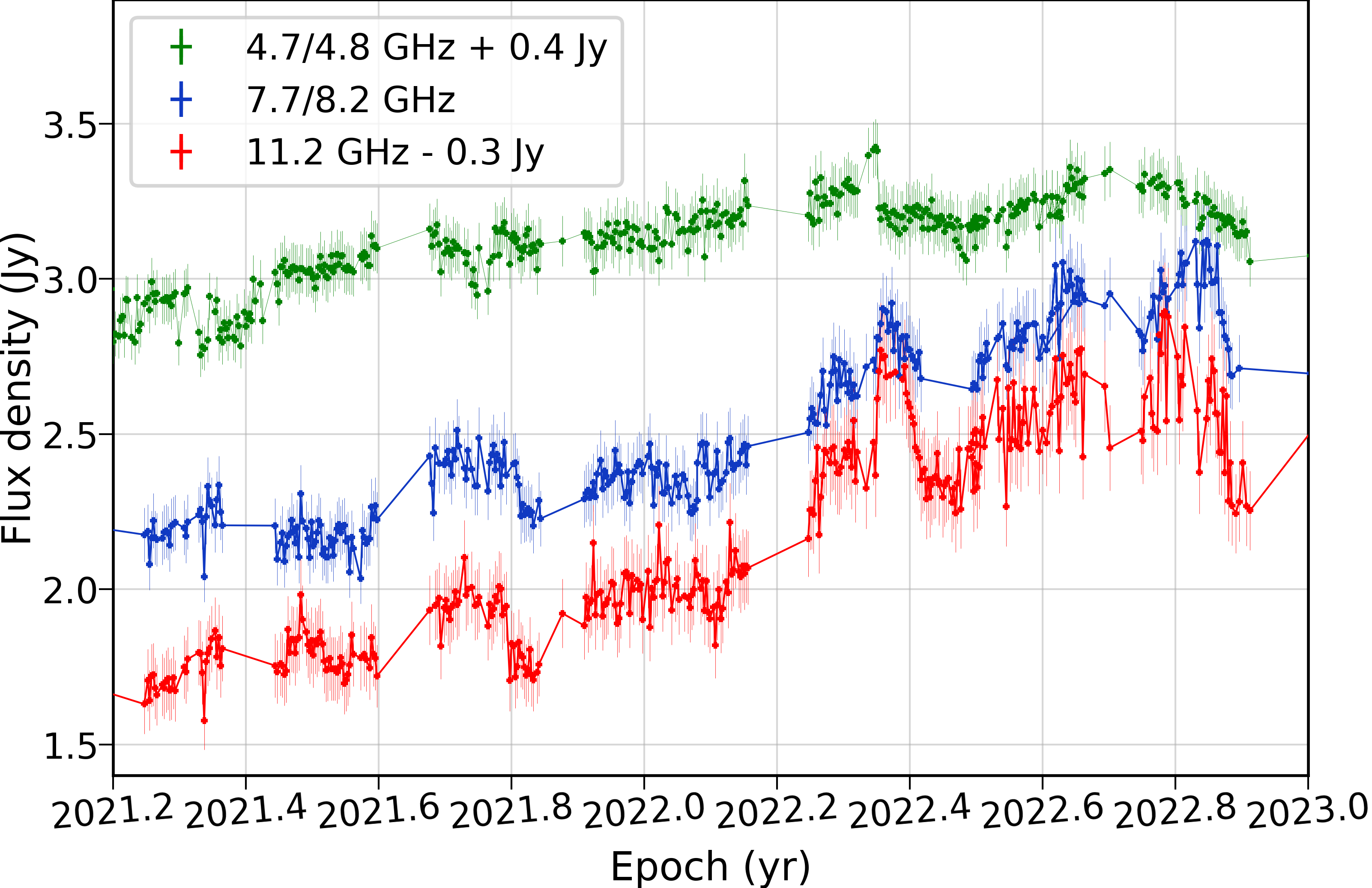}\vspace*{0.1cm} 
    \includegraphics[width=0.45\linewidth]{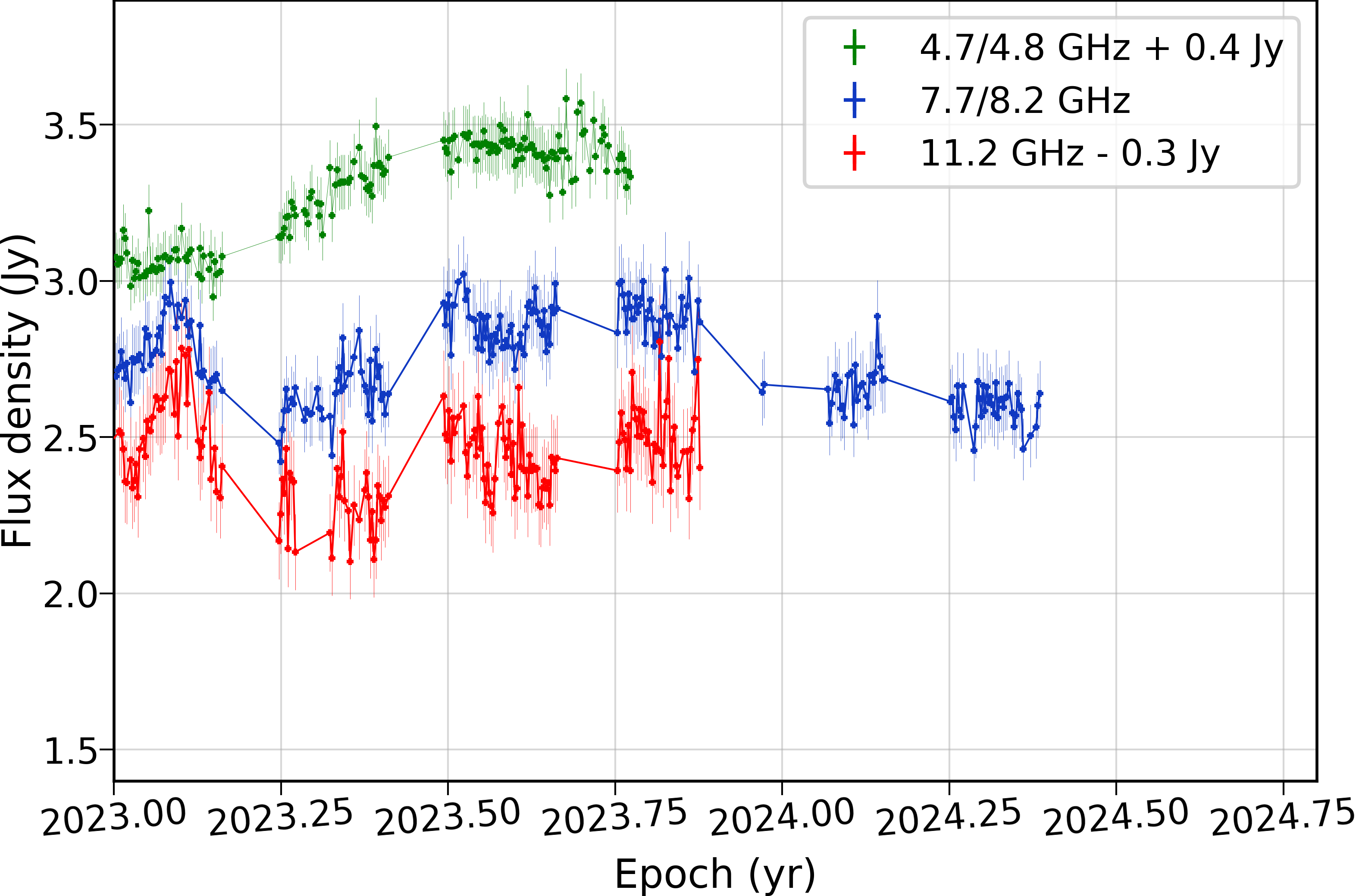}\vspace*{0cm} 
        
    \caption{The RATAN-600 light curves of 2005$+$403 measured at 4.7/4.8, 7.7/8.2, and 11.2~GHz (green, blue, and red colour, respectively). Yellow stars show the MOJAVE 15.4~GHz measurements. Some of the light curves have a flux density shift to avoid overlaying.}
    \label{fig:ratan_lcs_zoomed}
\end{figure*}

ESEs show great variety in form and amplitude of the event on light curves, but a common feature is a pair of peaks (or caustics) separated by weeks or months (local or distant screen), surrounding a flat or rounded minimum. A flat minimum occurs when the apparent size of the lens is larger than that of the source, while a rounded minimum is observed when the apparent size of the source exceeds that of the lens. Form variety is associated with the difference in the relative angular sizes between a background compact extragalactic source, scattering lens, and angular broadening of a source, as well as with the transverse velocity of the lens across the line of sight. In contrast to RISS, ESE is a rare event and, to date, it is known around 10 AGNs that demonstrated ESE on their light curves. The presence of both RISS and ESE effects with comparable amplitudes may complicate interpretation and search for possible ESEs. To better reveal short-term excursions in flux density changes we also show zoomed and sliced by 2 years the RATAN-600 light curves at 4.7/4.8, 7.7/8.2, and 11.2~GHz in \autoref{fig:ratan_lcs_zoomed}. 

According to the ESE theory, lensing should create symmetrical features on the light curve. However, it is quite difficult to find such symmetry in real data due to distortions on different time scales, e.g. 
\begin{enumerate}
    \item long-term trend associated with the internal evolution of the AGN (emergence and propagation down the jet of a new bright component of the outflow);
    \item the Earth’s orbit motion affects the projected speed of the scattering screen relative to the observer resulting in a specific asymmetry in the observed light curve features;
    \item time superposition of different propagation effects with comparable amplitudes (ESE, SRV, RISS, etc.);
    \item complex source or lens morphology;
    \item curvature in the lens proper motion.
\end{enumerate}

There is also the possibility that proper motion will cause the background source to pass near but not across the lens, resulting in a single broad maximum with no associated minimum \citep{Fiedler1994, Dong2018}. All of the features mentioned above complicate the identification and modeling of the ESE.

To automatically find the ESE candidates we calculated the maximum amplitude of flux density variations in a window of 2--7 months stepping over the light curve. To minimize the influence of RISS and long-term trend on the calculated variations of amplitudes, we subtracted a curve averaged over $\pm0.02$ years ($\pm1$ week) from the corresponding light curve. We visually inspected those periods which showed the highest variation amplitudes (relative variation exceeded 5~per~cent of the median) having more or less symmetrical shape, separate caustics (without overlapping with second ESE) and observed simultaneously at different frequencies. The lags between the light curves at different frequencies during the ESE periods measured by zDCF method do not exceed around $\pm5$~days. Such a small lag may occur just due to strong distortions (different for each frequency) of the ESE shape (imperfect ESE shape) caused by a number of effects discussed above, which highly influence the result of zDCF tests. The found periods can be potentially associated with ESE. This way, we identified:

\begin{enumerate}
\item the first event at epoch around June 2011, it clearly shows specific symmetric variations attributed to ESE simultaneously measured at all frequencies of 4.7/4.8, 7.7/8.2, and 11.2~GHz (\autoref{fig:ratan_lcs}, bottom left);

\item the quasar appears to have undergone another ESE started around March 2015. These modulations were detected only at 8.2 and 11.2~GHz (\autoref{fig:ratan_lcs}, bottom middle). In particular, at least six events after this epoch up to June 2016 also showed some ESE-like modulations following each other and partly overlapping in caustics. We discuss it in more detail in \autoref{s:multiple_ese};

\item the last event, at epoch around July 2020, shows simultaneous excursions at 4.7/4.8, 7.7/8.2, and 11.2~GHz (\autoref{fig:ratan_lcs}, bottom right). 
\end{enumerate}

\noindent
The duration of the events is approximately four months, and the maximum flux density variation reaches 10~per~cent at 11.2~GHz. It is worth noting that the amplitude of these variations increases with frequency. The lower the frequency, the higher the scattering power. We assume that this may be due to the fact that at low frequencies the source may already have too large observed size to produce a high-amplitude ESE and separate caustic surfaces. The extended background source smooths out caustic peaks. When the angular size of the observed radio source is much larger than the projected lens size, the result of refraction is ripple rather than strong symmetric modulations on the light curve \citep{Clegg_1998}. Also, the fraction of the flux density of the core component increases with frequency, causing a brighter effect at higher frequencies. Solar elongation of the source during the ESEs is large ranging from $76\degr$ to $103\degr$. Therefore, the formation of these modulations is unlikely to have been the result of SRV (scattering by the solar wind, \citealt{Marchili2024}) because of such a high separation from the Sun.

The RATAN 4.7/4.8~GHz light curve manifests a long-term trend, which is in good agreement with the trends observed at higher frequencies (7.7/8.2 and 11.2~GHz). However, unusual high-amplitude ($\approx20$ per~cent) flux density excursions were detected between 2010 and 2018. Notably, after 2018 these high-magnitude excursions virtually disappear. We constructed the Generalized Lomb-Scargle Periodogram (GLS) separately for the 4.7/4.8~GHz light curve before and after 2018. Originally, we de-trended the light curve using the boxcar technique over a time interval of $\pm0.2$ yrs to eliminate the influence of the long-term variability observed in the curve. GLS periodograms clearly revealed that the high-power variations with the time scale of $\sim1.36$~yr disappeared after 2018. The origin of these high-amplitude modulations before 2018 is still unclear and need to be further investigated elsewhere.

\section{Multi-frequency ESE fitting}
\label{s:ese_fitting}

The quasar 2005$+$403 clearly showed specific multi-frequency variations attributed to ESE around 2011.5, 2015.2, and 2020.6. For a quantitative description of the properties of interstellar plasma lenses we fitted the observed flux density variations from the RATAN-600 light curves at 4.7/4.8, 7.7/8.2, and 11.2~GHz using Fiedler’s statistical model of flux redistribution \citep{Fiedler1994} and Clegg's gaussian plasma lens model \citep{Clegg_1998}. The observed scattering events occurred at 2--3 frequencies simultaneously. This provides us an opportunity to fit them together, taking into account that: 

\begin{enumerate}
    \item the intrinsic size of the source scales with frequency as $\theta_{\rm s}~\propto~\nu^{-k_{\rm int}}$, where $k_{\rm int} = 1$ is the intrinsic power-law index. Here, we assume that the scattered component is the VLBI core~-- an apparent jet base of AGN, which observed size inversely proportional to $\nu$ for conical jet shape \citep{BK79}; 

    \item due to angular broadening the observed angular size of the source scales with frequency as $\theta_{\rm b}~\propto~\nu^{-k_{\rm scat}}$, where $k_{\rm scat}$ was derived in \cite{Koryukova2023} as $2.00\pm0.08$ which is in a good agreement with Gaussian screen model \citep{Cordes86}. 
\end{enumerate}

We also considered the possibility that the innermost optically thin component was scattered instead of the core implying that $k_{\rm int}~=~0$. However, this scenario results in too small fitted intrinsic size $\theta_{\rm int}$ at 1~GHz ($\approx 0$). We believe that the VLBI-core scattering assumption is more appropriate in this case and we use it in the multi-frequency fitting of ESE.

\subsection{Single ESE modeling results}
\label{s:single_ese}

\begin{figure*}
    \centering
    \includegraphics[width=1\linewidth]{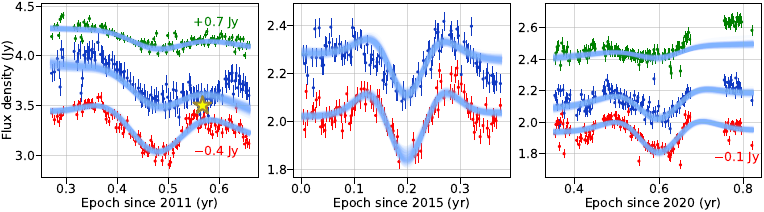}\vspace*{0cm}
    \caption{The RATAN-600 4.7/4.8, 7.7/8.2, and 11.2~GHz light curves (green, blue, and red colour, respectively) of the quasar 2005$+$403 attributed to ESE (green, blue, and red dots, respectively). Thin blue curves show the model predictions for 100 parameters values from the posterior distribution. Yellow star depict flux density measurement from the MOJAVE 15.4~GHz data that appeared to coincide with the right caustic of ESE. The total intensity map for this epoch shown in \autoref{fig:mojave_maps}, the model in \autoref{tab:mojave_model}. Error bars show fitted uncertainty at each frequency band (see \autoref{sec:fitting} for the details).} 
    \label{fig:ese_fiedler_fitting_result}
\end{figure*}

The quasar appears to have undergone an ESE started around May 2011. Characteristic modulations are observed at all the RATAN-600 observation frequencies, i.e. 4.8, 7.7, and 11.2~GHz. The maximum flux density variation reaches 10~per~cent at 11.2~GHz. The results of fitting are illustrated in \autoref{fig:ese_fiedler_fitting_result} (left), estimated model parameters are listed in \autoref{tab:model_parameters_fiedler}. The posterior distribution of the model parameters for this ESE is depicted in the Appendix (\autoref{fig:posterior_ese_2011}). The reconstructed lens and source parameters are listed in \autoref{tab:lens_parameters_fiedler}  (third column).

{\renewcommand{\arraystretch}{1.5}%
\begin{table*}
    \caption{Fiedler's and Clegg's model parameters estimated for ESE observed on the RATAN-600 light curves at different epochs.}
    \centering
    \begin{tabular}{|*{5}{c|}}
    \hline    
        Parameter             &Prior bounds& \multicolumn{3}{c}{Median and 1$\sigma$-uncertainty intervals} \\
        \hline
                              && 2011      & 2015           & 2020  \\\cline{3-5}
         $\Delta t$               &&  $0.39$   & $0.37$         & $0.47$ \\
         $\mu_{\rm s}$        &$0.1-20.0$  &  $8.14^{+0.40}_{-0.32}$   & $13.24^{+1.50}_{-1.02}$   & $8.45^{+0.53}_{-0.39}$\\
         $\theta_{\rm l,s,1}$ &$0.1-5.0$   & $0.34^{+0.06}_{-0.06}$    & $0.71^{+0.20}_{-0.21}$    & $0.19^{+0.08}_{-0.05}$\\ 
         $\theta_{\rm b,s,1}$ &$0.1-5.0$   & $1.54^{+0.19}_{-0.11}$    & $1.16^{+0.29}_{-0.17}$    & $2.07^{+0.49}_{-0.36}$\\ 
         $f_{\rm core1}$      &$0.0-1.0$   & $0.20^{+0.04}_{-0.04}$    & $0.42^{+0.12}_{-0.11}$    & $0.18^{+0.07}_{-0.06}$\\
         $f_{\rm core2}$      &$0.0-1.0$   & $0.39^{+0.08}_{-0.08}$    & $0.75^{+0.18}_{-0.22}$    & $0.56^{+0.13}_{-0.16}$\\
         $f_{\rm core3}$      &$0.0-1.0$   & $0.89^{+0.08}_{-0.15}$    & \ldots                    & $0.81^{+0.14}_{-0.22}$\\
         $\delta$             &$-0.1-0.1$  & $-0.012^{+0.002}_{-0.002}$& $-0.009^{+0.001}_{-0.001}$& $-0.013^{+0.002}_{-0.002}$\\
         $m_1$                &$-1.0 - 0.5$& $-0.017^{+0.002}_{-0.002}$& $0.003^{+0.002}_{-0.002}$ & $0.009^{+0.002}_{-0.002}$\\
         $m_2$                &$-1.0 - 0.5$& $-0.041^{+0.005}_{-0.005}$& $0.001^{+0.002}_{-0.001}$ & $0.011^{+0.002}_{-0.002}$\\
         $m_3$                &$-1.0 - 0.5$& $-0.021^{+0.002}_{-0.002}$& \ldots                    & $0.001^{+0.002}_{-0.002}$\\[1ex]\cline{1-5}
         $\alpha$             &$0.1 - 10.0$& $0.68^{+0.50}_{-0.50}$     & $0.22^{+0.03}_{-0.02}$    & $1.41^{+0.21}_{-0.19}$  \\
    \hline
\end{tabular}
    \begin{tablenotes}
        \item Note: Index '1' in parameter name means that this value was estimated for the lowest frequency data used for fitting. Other indices correspond to higher frequencies in the set in ascending order. For ESE in 2015 we used only two frequencies for fitting (8.2 and 11.2~GHz).
        \item The parameters are as follows: $\Delta t$~-- the duration of the event in yr, $\mu_{\rm s}$~-- proper motion of the lens in $\theta_{\rm s}$ units, $\theta_{\rm l,s,1}$~-- lens angular size in $\theta_{\rm s}$ units, $\theta_{\rm b,s,1}$~-- broadening angle of the source in $\theta_{\rm s}$ units, $f_{\rm core1}$, $f_{\rm core2}$, $f_{\rm core3}$~-- core dominance of the flux density that was scattered at different frequencies, $\delta$~-- the epoch of ESE minimum shift measured in mas, $m_1$, $m_2$, $m_3$~-- slope of a light curve fitted with a linear trend separately for each frequency, $\alpha$~-- lens strength fitted at 11.2~GHz using Clegg's model.
        %[1ex]
    \end{tablenotes}
\label{tab:model_parameters_fiedler}
\end{table*}}

{\renewcommand{\arraystretch}{1.7}%
\begin{table*}
    \caption{The summary of plasma lens and quasar 2005$+$403 properties derived from ESEs fitting at three different epochs using Fiedler's and Clegg's models.}
    \centering
    \begin{tabular}{llcccccl}
    \hline    
        Frequency & Parameter         & Median $\pm1\sigma$    && Median $\pm1\sigma$    && Median $\pm1\sigma$              & Unit\\
        (1)       &(2)                &(3)                     && (4)                    && (5)                              & (6)\\
        \hline
                  &                   & 2011.48                &&2015.20                 && 2020.60                          &    \\\cline{3-7}

                  &$\theta_{\rm l}$   & 0.41$\pm 0.08$         && 0.38$\pm 0.06$         && 0.17$\pm 0.04$         & mas  \\ 
                  &$a_{\rm l}$        & 0.73$\pm 0.15$         && 0.68$\pm 0.11$         && 0.30$\pm 0.06$         & au  \\
                  &$\mu_{\rm l}$      & 9.88$\pm 1.98$         && 7.19$\pm 1.18$         && 7.71$\pm 1.64$         & mas yr$^{-1}$  \\ 
                  &$V_{\rm l}$        & 83.8$\pm 16.8$ $^a$    && 61.1$\pm 10.0$ $^b$    && 65.4$\pm 14.1$ $^c$    & km s$^{-1}$  \\ \cline{2-8}
                  &$n_{\rm e}$        & $1439^{+109}_{-118}$   && $913^{+123}_{-107}$    && $1199^{+185}_{-158}$   & cm$^{-3}$ \\
                  &$M_{\rm l}$              & $0.42^{+0.03}_{-0.03}$ && $1.89^{+0.24}_{-0.21}$ && $0.02^{+0.01}_{-0.01}$ & $10^{-15}\rm M_\odot$ \\
         \hline
     4.7/4.8 GHz  &$\theta_{\rm b}^*$ & 2.73$^{+0.24}_{-0.22}$ && \ldots                 && 2.84$^{+0.24}_{-0.22}$ & mas  \\ 
                  &$\theta_{\rm s}$   & 1.77$^{+0.13}_{-0.19}$ && \ldots                 && 1.37$^{+0.29}_{-0.26}$ & mas  \\ 
                  &$S_{\rm scat}$     & 0.68$^{+0.14}_{-0.12}$ && \ldots                 && 0.43$^{+0.16}_{-0.14}$ & Jy  \\ 
                  &$S_{\rm unscat}$   & 2.78$^{+0.12}_{-0.14}$ && \ldots                 && 2.02$^{+0.14}_{-0.16}$ & Jy  \\
     7.7/8.2 GHz  &$\theta_{\rm b}^*$ & 1.64$^{+0.15}_{-0.14}$ && 1.09$^{+0.17}_{-0.16}$ && 1.63$^{+0.14}_{-0.13}$ & mas  \\ 
                  &$\theta_{\rm s}$   & 1.11$^{+0.08}_{-0.12}$ && 0.94$^{+0.16}_{-0.19}$ && 0.79$^{+0.16}_{-0.15}$ & mas  \\ 
                  &$S_{\rm scat}$     & 1.45$^{+0.29}_{-0.28}$ && 0.94$^{+0.26}_{-0.26}$ && 1.19$^{+0.28}_{-0.33}$ & Jy  \\ 
                  &$S_{\rm unscat}$   & 2.22$^{+0.28}_{-0.29}$ && 1.33$^{+0.26}_{-0.26}$ && 0.95$^{+0.33}_{-0.28}$ & Jy  \\
        11.2 GHz  &$\theta_{\rm b}^*$ & 1.17$^{+0.10}_{-0.09}$ && 0.80$^{+0.13}_{-0.12}$ && 1.19$^{+0.09}_{-0.09}$ & mas  \\ 
                  &$\theta_{\rm s}$   & 0.76$^{+0.06}_{-0.08}$ && 0.69$^{+0.12}_{-0.14}$ && 0.58$^{+0.12}_{-0.11}$ & mas  \\ 
                  &$S_{\rm scat}$     & 3.31$^{+0.29}_{-0.54}$ && 1.53$^{+0.36}_{-0.44}$ && 1.62$^{+0.29}_{-0.46}$ & Jy  \\ 
                  &$S_{\rm unscat}$   & 0.39$^{+0.54}_{-0.29}$ && 0.49$^{+0.44}_{-0.36}$ && 0.42$^{+0.46}_{-0.29}$ & Jy  \\ 
    \hline
\end{tabular}
    \begin{tablenotes}
        \item The columns are as follows: (1) central observing frequency; (2) name of the parameter; (3), (4), (5) estimated value of the parameter for ESE with the epoch of minimum at 2011.48, 2015.20, and 2020.60, respectively, with the range of uncertainty on a $1\sigma$ level; (6) unit.
        \item The parameters are as follows: $\theta_{\rm l}$ and $a_{\rm l}$ are the lens angular and linear size, $\mu_{\rm l}$ is the proper motion of the lens, $V_{\rm l}$ is the transverse venosity of the lens, $n_{\rm e}$ is the maximum free-electron density of the lens, $M_{\rm l}$ is the mass of the lens, $\theta_{\rm b}$ is the angular broadening of the source, $\theta_{\rm s}$ is the intrinsic size of the source, $S_{\rm scat}$ and $S_{\rm unscat}$ are lensed and unlensed parts of the nominal flux density level.
        \item $^{\rm a,b,c}$ The transverse velocity corrections for the Earth's motion around the Sun are $-9.8$, $+18.2$, and $-7.1$~km s$^{-1}$ for $V_{\rm l}$ calculated in 2011.48, 2015.20, and 2020.60, respectively.
        \item $^{*}$ This parameter was calculated according to \autoref{eq:observed_size}.
        \item Note 1: ESE in 2015.20 detected only at two frequencies.
        \item Note 2: $n_{\rm e}$ and $M_{\rm l}$ were fitted applying Clegg's model to the data at 11.2~GHz. 
    \end{tablenotes}
\label{tab:lens_parameters_fiedler}
\end{table*}}

At 7.7~GHz, the right caustic shows an atypical distortion about the epoch of late June 2011. One of the observing epoch of 2005$+$403 in MOJAVE 15.4~GHz VLBA\footnote{\url{https://www.cv.nrao.edu/MOJAVE/sourcepages/2005+403.shtml}} project serendipitously coincides with this distortion in time, therefore we carefully analyzed the brightness distribution and performed structure model fitting for this epoch. In \autoref{fig:mojave_maps}, we show the zoomed brightness distribution map of the quasar which depicts a characteristic for 15.4~GHz morphology with the jet developing in $\mathrm{PA} = 93\degr$ within first 2~mas, then the jet changes its direction to south-east, with $\mathrm{PA} = 126\degr$ \citep{Koryukova2023}. The best-fitting model contains five Gaussian components, where the brightest feature is not the core, but the optically thin component (J1 in \autoref{fig:mojave_maps}). We suppose that these two compact and bright features in the jet's origin can significantly influence the resulting ESE form and symmetry, because the screen may scatter both of these bright and compact components creating peculiar light curve shapes. Moreover, we found that at this epoch both the core and the jet components have secondary images due to anisotropic scattering in the interstellar medium. The scatter-induced patterns are extended in a direction close to the line of constant Galactic latitude ($\mathrm{b} = 4\fdg3$) at $\mathrm{PA} = 40\degr$.

The second ESE started around February 2015 is detected at 8.2 and 11.2~GHz. The results of the fitting are illustrated in \autoref{fig:ese_fiedler_fitting_result} (middle); estimated model parameters are listed in \autoref{tab:model_parameters_fiedler}. The posterior distributions of the model parameters for these ESE are depicted in \autoref{fig:posterior_ese_2015}. The reconstructed lens and source parameters are given in \autoref{tab:lens_parameters_fiedler} (fourth column). Finally, the last found ESE started around 2020 June observed at all frequencies, i.e. 4.7, 8.2, and 11.2~GHz. The results of the fitting are illustrated in \autoref{fig:ese_fiedler_fitting_result} (right); and the estimated model parameters are listed in \autoref{tab:model_parameters_fiedler}. The posterior distribution of the model parameters for these ESE is depicted in \autoref{fig:posterior_ese_2020}. The reconstructed lens and source parameters are listed in \autoref{tab:lens_parameters_fiedler} (fifth column). The fitting results show that $f_{\rm core}$ parameter increases with frequency. It is consistent with our assumption for the scattering of the VLBI core component. The fraction of the total flux density that was scattered increases with frequency due to steep spectrum of the optically-thin jet emission and flat spectrum of the core.

ESE models operate with relative sizes (lens to source or source to lens size). Therefore the reconstructed lens parameters turned out to be frequency-dependent, since the intrinsic size of the VLBI core at different frequencies is different. As a final result we averaged the lens parameters calculated using different frequencies per epoch. According to the model, a plasma lens typically covers the source for about four months. The estimated angular size of a scattering lens ($\theta_{\rm l}$) in 2011, 2015, and 2020 was lower than the source's angular size ($\theta_{\rm s}$) resulting in rounded-minimum ESE shapes. The linear lens sizes were estimated as $a_{\rm l} = \theta_{\rm l}D$ with the average over all ESEs $a_{\rm l} = 0.6\pm0.1$~au. The average proper motion of plasma lens across the line of sight is $\mu_{\rm l} = 8.3\pm0.7$~mas~yr$^{-1}$ which corresponds to the transverse velocity of the lens with respect to the observer $V_{\rm l} = 4.74\,\mu_{\rm l} \times D \approx 70.1\pm5.7$~km~s$^{-1}$, where $D = 1.79\pm0.10$~kpc is the assumed distance to the lens \citep{Rygl2012}. \cite{Rygl2012} measured the distances to the five massive star-forming regions towards the Cygnus~X complex within a 10~per~cent accuracy. The line of sight to the source also passes through the Perseus Arm ($\sim 6$~kpc) and the Outer Arm ($\sim 10$~kpc) regions, which may also potentially host a scattering screen. However, in this case, the inferred transverse velocity of the lens would exceed the velocity on the Galaxy rotation curve. Thus, most likely, the screen is located in the Local Arm, no further than the star-forming regions in Cygnus~X with the average distance of 1.79~kpc.

In addition to the parameters of lenses mentioned above, the Clegg's model allowed us to make a rough (due to strong degeneracy of the parameters) estimation of maximum free-electron density on the line of sight ($n_{\rm e}$) and mass of the lens ($\rm M_\odot$). We fitted only the light curves at the highest frequency (11.2~GHz) due to the highest amplitude of ESE feature. We obtained a-priori information about the parameters from the results of fitting with Fiedler's model at 11.2~GHz, i.e. measured source-to-lens ratio ($\theta_{\rm s}/\theta_{\rm l}$), proper motion ($\mu$), core dominance ($f_{\rm core3}$), horizontal shift ($\delta$), slope ($m_{\rm 3}$) and fitted only the lens strength ($\alpha$) for each ESE and reconstructed $n_{\rm e}$ and $\rm M_\odot$ according to \autoref{eq:Clegg_alpha} and other equations mentioned in \autoref{s:Clegg_model}. The obtained values of $\alpha$ are listed in \autoref{tab:model_parameters_fiedler}. Also, we show the posterior distribution of $\alpha$ for each ESE in the Appendix (\autoref{fig:alphas}). The estimated values of $n_{\rm e}$ and $\rm M_\odot$ are listed in \autoref{tab:lens_parameters_fiedler}.

\subsection{Multiple ESE observed in 2015.1--2016.5}
\label{s:multiple_ese}

\begin{figure*}
    \centering
    \includegraphics[width=0.8\linewidth]{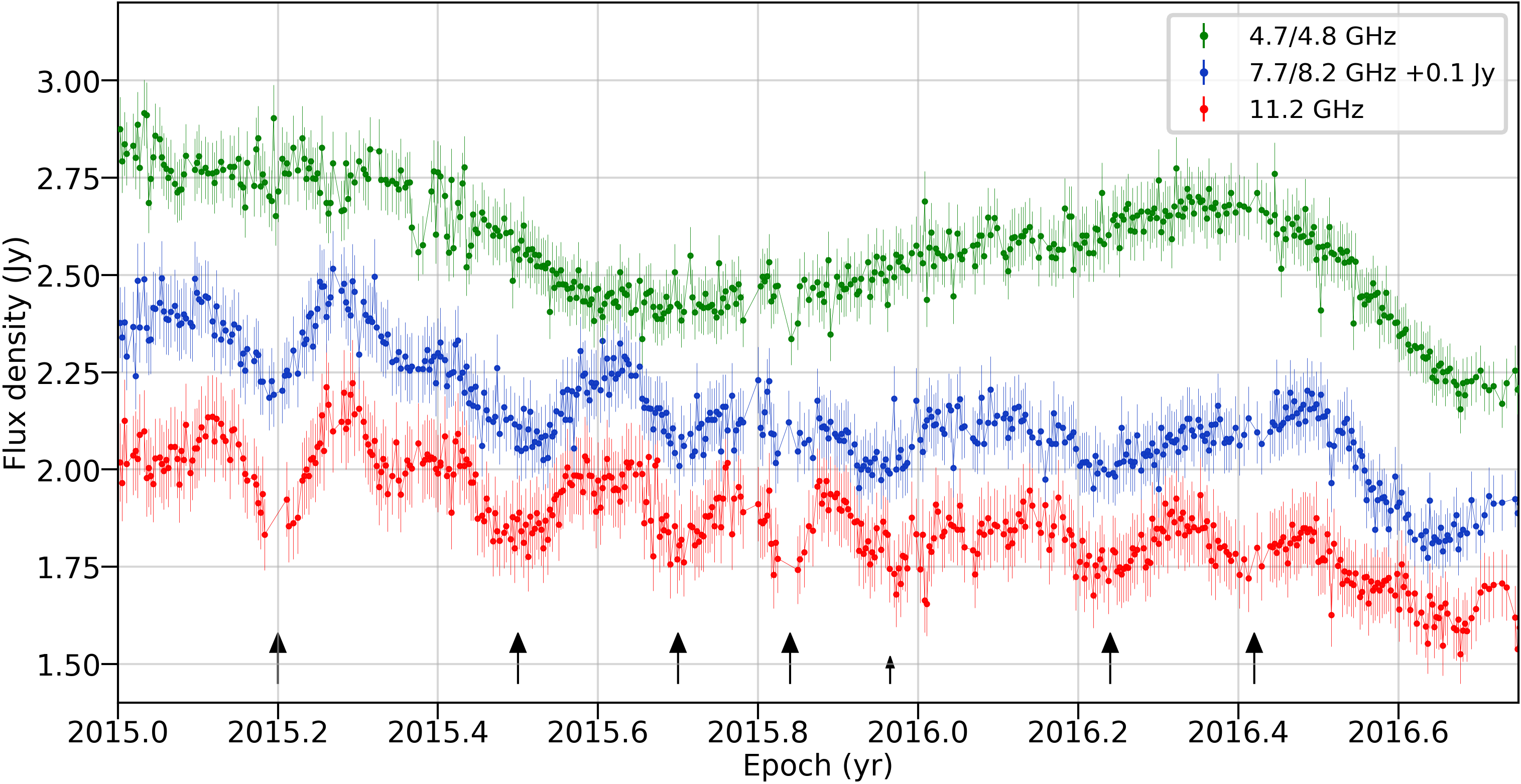}\vspace*{0cm}
    \caption{The RATAN-600 4.7/4.8, 7.7/8.2, and 11.2~GHz light curves (green, blue, and red curves, respectively) that show multiple-ESE in a row. At least six ESE events are observed (black arrows) and one candidate that highlighted by small arrow.}
    %gray
    \label{fig:multiple_ese}
\end{figure*}

The flux density variations of the quasar 2005$+$403 between 2015.1 and 2016.6 appear as a regular sequence of ESE-like modulations. Up to six ESE features are observed in a row (black arrows in \autoref{fig:multiple_ese}) and one additional candidate (marked with gray arrow) that looks like ripples. When the observed angular size of the background radio source is much larger than the projected size of the lens, the result of refraction is ripples rather than strong symmetric modulations in the light curve \citep{Clegg_1998}. The average cadence of the events is 2.5-months (\autoref{fig:multiple_ese}), with the duration of a single event is of the order of several months. These modulations generally affect both 7.7/8.2 and 11.2~GHz light curves, but not 4.7/4.8~GHz.

We tested the SRV scenario comparing the epochs of minimums of the observed variations with epochs of minimum and maximum solar elongation. Apparently, these series of modulations were not significantly affected by solar elongation, as was shown for the source 0537$-$158 in \citealt{Marchili2024} (see fig.~4). Only one low-amplitude modulation (at 2016.4) that is observed only at 11.2~GHz coincides with minimum solar elongation.

It was theoretically predicted that the multi-component structure of the background source, as well as the clumpy inhomogeneous structure of the scattering screen are able to induce more than one ESE in a row on the light curve \citep{Fiedler1994}. We assume that the observed sequence of six modulations on the 2005$+$403 light curves may be caused by scattering effects. To our knowledge, it is the first reported case of multiple ESEs occurred in a row. It is noteworthy that the events vary in shape, amplitude, and duration, indicating that the properties of the scattering screens, i.e. free-electron density, distance, linear size or transverse velocity, differ significantly. Caustics at 4.7/4.8~GHz are not noticeable and may be smoothed down due to stronger angular broadening of the source (\autoref{fig:multiple_ese}, green curve).

In \autoref{fig:mojave_maps} we show the first two components of the model that are located at the jet origin and are subject to scattering. We used elliptical Gaussian to fit these components. The complete results of model fitting are presented in \autoref{tab:mojave_model}. The brightest component (J1 in \autoref{fig:mojave_maps}) is the innermost optically thin feature of the jet. The PA of the major axis of the ellipse ($37.7\degr$) is close to the $\rm PA = 40\fdg6$ of the constant Galactic latitude line ($b= \rm const$, yellow dotted line in \autoref{fig:mojave_maps}). Apparently, the presence of refractive-dominated scattering formed the secondary image of this component and stretched it along the direction of $b= \rm const$. The formation of sub-images along this direction may indicate a flat geometry of the screen, oriented perpendicular to the Galactic plane. If the screen is quasi-spherical or has any other curved geometry, then the sub-image may be formed in any other direction (not $b= \rm const$), depending on the direction of screen motion across the line of sight.

The rightmost component (C) is also subject to anisotropic scattering. The PA of the major axis of the ellipse is oriented at $29.3\degr$ which is close to the PA of $b= \rm const$ line. The core (C) and the innermost bright component of the jet (J1) appear to be very close in projection to the observer ($0.45$~mas), since these components propagate at a very small angle relative to the line of sight (as shown in \autoref{fig:mojave_maps} and discussed in \citealt{Gabani2006}). The presence of several bright and compact components in the jet origin of AGN, as well as multiple screens crossed the line of sight one by one in projection and located at different distances from the Earth or a combination of these two effects may induce multiple ESE sequence on the light curve. The angular distance between the core and the innermost jet feature in our model (\autoref{tab:mojave_model}) projected to the line of constant Galactic latitude is $\approx 0.17$~mas. Multiplying this distance by the observed cadence of $\approx0.2$~yr we would obtain $\mu\approx0.85$~mas~yr$^{-1}$ which is much less than the characteristic $\mu$ estimated from modeling (\autoref{tab:lens_parameters_fiedler}). Thus, we can conclude that this scenario can not explain any the observed pairs of ESEs.

\subsection{Discussion}

The observed discrepancies between the modeled and observed light curves may result from numerous reasons or their combination, e.g. substructure within the lens, an anisotropic lens shape, a lens which only grazes the source rather than passing completely over it, or unresolved substructure within the extragalactic sources \citep{Clegg_1998}. The details about the geometry and physical origin of plasma lenses that cause ESE still remain a highly debatable question. The general ISM is widely believed to have an extended distribution of turbulent electron density fluctuations \citep{Rickett1990}. The energetic processes that stir the ISM (e.g. supernova explosions, ionization fronts, density waves in the spiral arms of the Galaxy, etc.) are typically short-lived and occur on parsec scales or larger. Whatever physical process produces the scattering lenses, they must persist on a timescale of months to years in order to match the timescale of ESE observations. According to the results of modeling, namely the linear sizes of lenses, proper motions of plasma lens across the line of sight, transverse velocities of the lenses with respect to the observer, etc., we suppose that radio filaments oriented predominantly perpendicular to the Galactic plane and probably related to pulsars (e.g. \citealt{Churazov2024}), or very straight and large-aspect-ratio, aligned density sheets with thicknesses $\approx$~au may be a cause of the observed ESEs \citep{Kempski2024}. Another possible explanation could be shock wave fronts of supernova remnants (e.g. \citealt{Romani1987}), though no supernova remnants have been observed in the sky region near 2005$+$403 yet. ESE requires high free-electron densities along the line of sight ($n_{\rm e} \geq 10^3\,\rm cm^{-3}$), which is not observed in supernova remnants. Other scenarios and geometries of scattering screens face the overpressure problem \citep{Stanimirovic2018}.

Note that the estimated properties of the plasma lenses that created the ESEs are in good agreement with each other, reflecting abundance of such structures for a given direction of the Galaxy. The ESEs occur quite frequently on the light curves of the quasar 2005$+$403 suggesting strong turbulence and heterogeneity of the plasma on the line of sight of the source. It was assumed that ESE occurs rarely, but for this source, this statement becomes doubtful.

Estimates of the intrinsic, unscattered angular size of the quasar ($\theta_s$) at the epochs of the ESE reconstructed assuming a conical jet shape \citep{BK79}, i.e. the dependence of the observed angular size of the source on the frequency is $\propto\nu^{-1}$. The obtained results are listed in \autoref{tab:lens_parameters_fiedler}.

\section{Summary}
\label{s:summary}

The extragalactic radio source 2005$+$403 observed towards the Cygnus region with low Galactic latitude showed great potential for probing the interstellar plasma and detecting various scattering effects, especially angular broadening of the size, multiple imaging effect, and ESEs on light curves. 

We investigated the RATAN-600 light curves of the source measured at 4.7/4.8, 7.7/8.2, and 11.2~GHz to search for evidence of radio waves scattering. The full RATAN-600 data sets contain daily observations that span more than 20 years. These light curves show a frequency-dependent time delay caused by synchrotron opacity: zDCF correlation analysis reveals that variations at 11.2~GHz occur about 2~months earlier than at 7.7/8.2~GHz, and about 6~months earlier than at 4.7/4.8~GHz. The time lag of 3.7~months corresponds to 7.7/8.2~GHz light curve variations leading variations at 4.7/4.8~GHz.

One of the VLBA observing epochs of the source, June 24, 2011, taken within the MOJAVE program at 15~GHz serendipitously coincided with the right caustic of an extreme scattering event detected on the RATAN-600 light curve suggesting a formation of secondary image(s) of a bright and compact feature of a background source. At this particular epoch the brightest feature was the innermost jet component, not the core. By structure modeling of the observed brightness distribution, we found quite convincing evidence of the refractive-dominated scattering occurring in both the core and innermost jet feature. The position angle of the major axis of the elliptical Gaussian components fitted to the data is closely oriented to the direction of the constant Galactic latitude line ($\rm PA = 40\fdg6$), while the jet direction is roughly orthogonal to it. We note that this is the first case of anisotropic scattering detected simultaneously in two separate VLBI components of the source structure that are bright and compact enough to create an edge effect leading to the formation of secondary images.

The RATAN-600 light curves are found to be characterized by variability on a wide range of time scales (days to years). The longest scale is about seven years and is clearly observed at all frequencies. The origin of the long-term variability may be caused by changing Doppler boosting, e.g. by jet precession. Modulations observed over shorter time intervals (1.5~years or less) most likely relate to propagation effects. Unusual high-amplitude flux density excursions with a time scale of 1.36~yr were detected only at the RATAN 4.7/4.8~GHz light curve before 2018, which virtually disappeared after 2018. The origin of these modulations is still unclear and need to be further investigated elsewhere. 

After an automatic search for ESE candidates, we visually inspected those periods which showed the highest variation amplitudes and identified three separate multi-frequency ESE modulations appeared in 2011, 2015, and 2020. The duration of the events was approximately four months, and the maximum flux density variations reach 10~per~cent at 11.2~GHz. The absence of a significant time lag between the modulations at different frequencies supports their scattering origin. None of the events coincide with the time of minimum solar elongation, which argues against scattering by solar wind. We estimated the following average lens parameters: the duration of the events ($\sim4$~months), angular and linear size of the scattering lenses $0.3\pm0.1$~mas and $0.6\pm0.1$~au, respectively, proper motion $8.3\pm0.7$~mas~yr$^{-1}$ and transverse velocity of lenses $70.1\pm5.7$~km~s$^{-1}$, maximum free-electron density on the line of sight $1183\pm124$~cm$^{-3}$ and lens mass $(0.8\pm0.4)\times10^{-15}\,\rm M_\odot$.  

%individually

We suppose that radio filaments oriented predominantly perpendicular to the Galactic plane and probably related to pulsars, or very straight and large-aspect-ratio density sheets with thicknesses $\approx$~au may be the cause of the observed ESEs. Another possible explanation could be shock wave fronts of supernova remnants, although this scenario faces difficulties in explaining the preference of the emission induced by scattering to be stretched out along the line of constant Galactic latitude.

The flux density variations at 7.7/8.2 and 11.2~GHz of the quasar 2005$+$403 between 2015.1 and 2016.6 appear as a regular sequence of up to six ESE-like modulations in a row. The average cadence of the events is 2.5-months, and the duration of a single event is of the order of several months. Only one event coincides with the epoch of minimum solar elongation, which may indicate a non-interstellar scattering origin. We suggest that the other five events appear due to scattering in the ISM. Both scenarios of multi-component structure of the background source as well as the clumpy inhomogeneous structure of the scattering screen are able to induce more than one ESE in a row on the light curve. We believe that the latter scenario is more likely than the first one. These modulations are not detected at 4.7/4.8~GHz and may be smoothed down due to the larger source-to-lens size ratio at low frequency. 

New techniques and new observations from current and next-generation radio telescopes will allow us to test different models that have been proposed to explain ESEs and shed light on the nature of the scattering screens.

\section*{Acknowledgements}
%We thank the anonymous referee for useful comments, which helped to improve the manuscript.

The MOJAVE database maintained by the MOJAVE team \citep{2018ApJS..234...12L}. This work made use of the Swinburne University of Technology software correlator, developed as part of the Australian Major National Research Facilities Programme and operated under licence.
This study was supported by the Russian Science Foundation grant\footnote{Information about the project: \url{https://rscf.ru/en/project/25-22-00152/}} 25-22-00152.

%%%%%%%%%%%%%%%%%%%%%%%%%%%%%%%%%%%%%%%%%%%%%%%%%%
\section*{Data Availability}

The RATAN-600 light curves underlying this article will be shared on reasonable request to S.A. Trushkin (sergei.trushkin@gmail.com).
%if we use VLBA:
In this study, we used the data from the MOJAVE sessions. The original data is available at the NRAO \href{https://data.nrao.edu/}{website}.

%%%%%%%%%%%%%%%%%%%% REFERENCES %%%%%%%%%%%%%%%%%%

\bibliographystyle{mnras}
\bibliography{article}

%%%%%%%%%%%%%%%%%%%%%%%%%%%%%%%%%%%%%%%%%%%%%%%%%%

\appendix
\section{Posterior distributions from fittings}
\label{s:appendix}

Here, we present all the posterior distributions of the estimated model parameters  as a result of the ESE modeling using Fiedler's and Clegg's model.

\begin{figure*}
    \centering
    \includegraphics[width=1\linewidth]{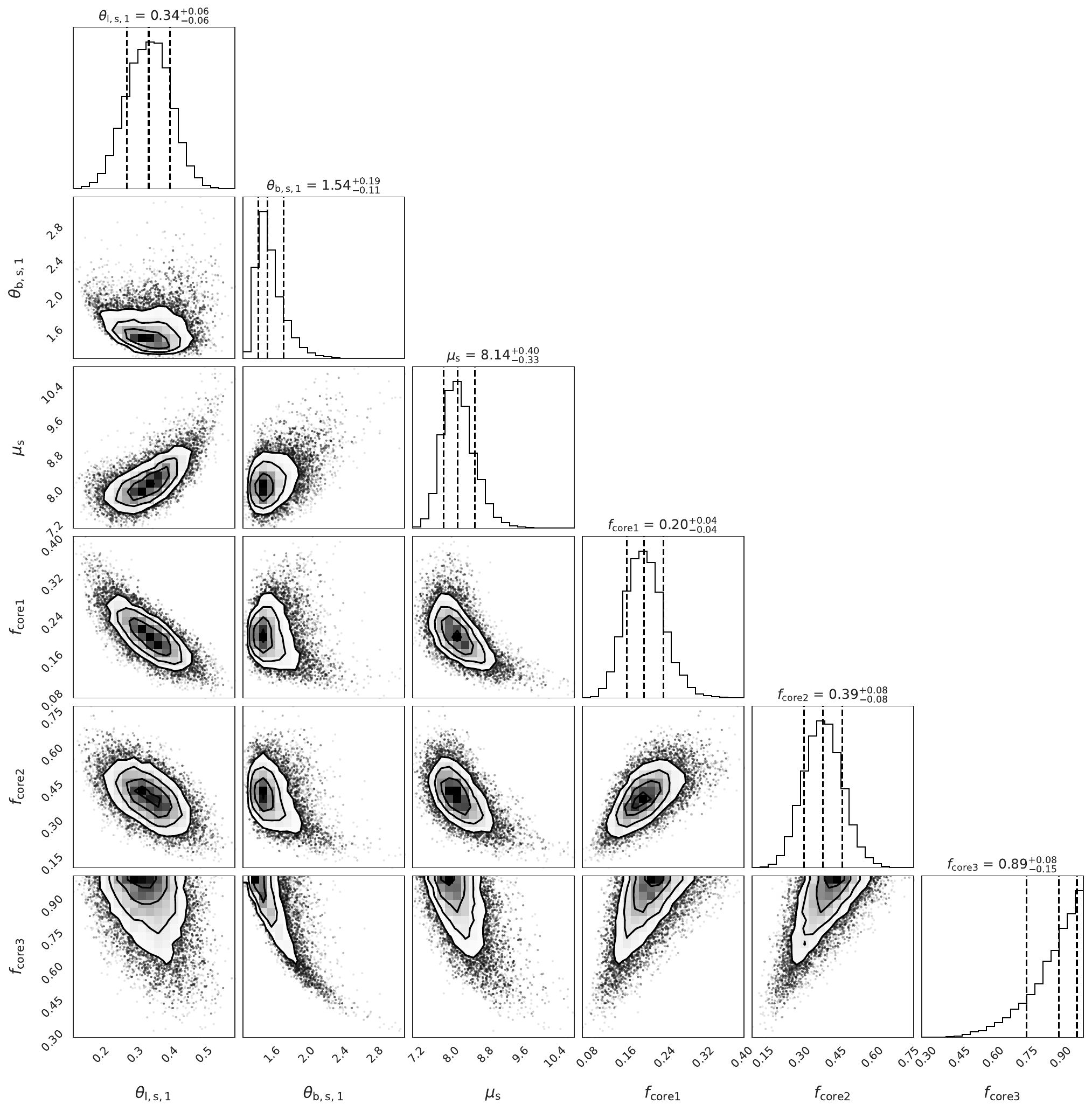}
    \caption{Posterior distribution of the estimated model parameters for the ESE in 2011 using Fiedler's model. Index '1' in parameter name means that this value was estimated for the lowest frequency, i.e. 4.8~GHz. Other indices correspond to higher frequencies in the set in ascending order. Dashed lines represent median and $\pm 1\sigma$.}
    \label{fig:posterior_ese_2011}
\end{figure*}

\begin{figure*}
    \centering
    \includegraphics[width=1\linewidth]{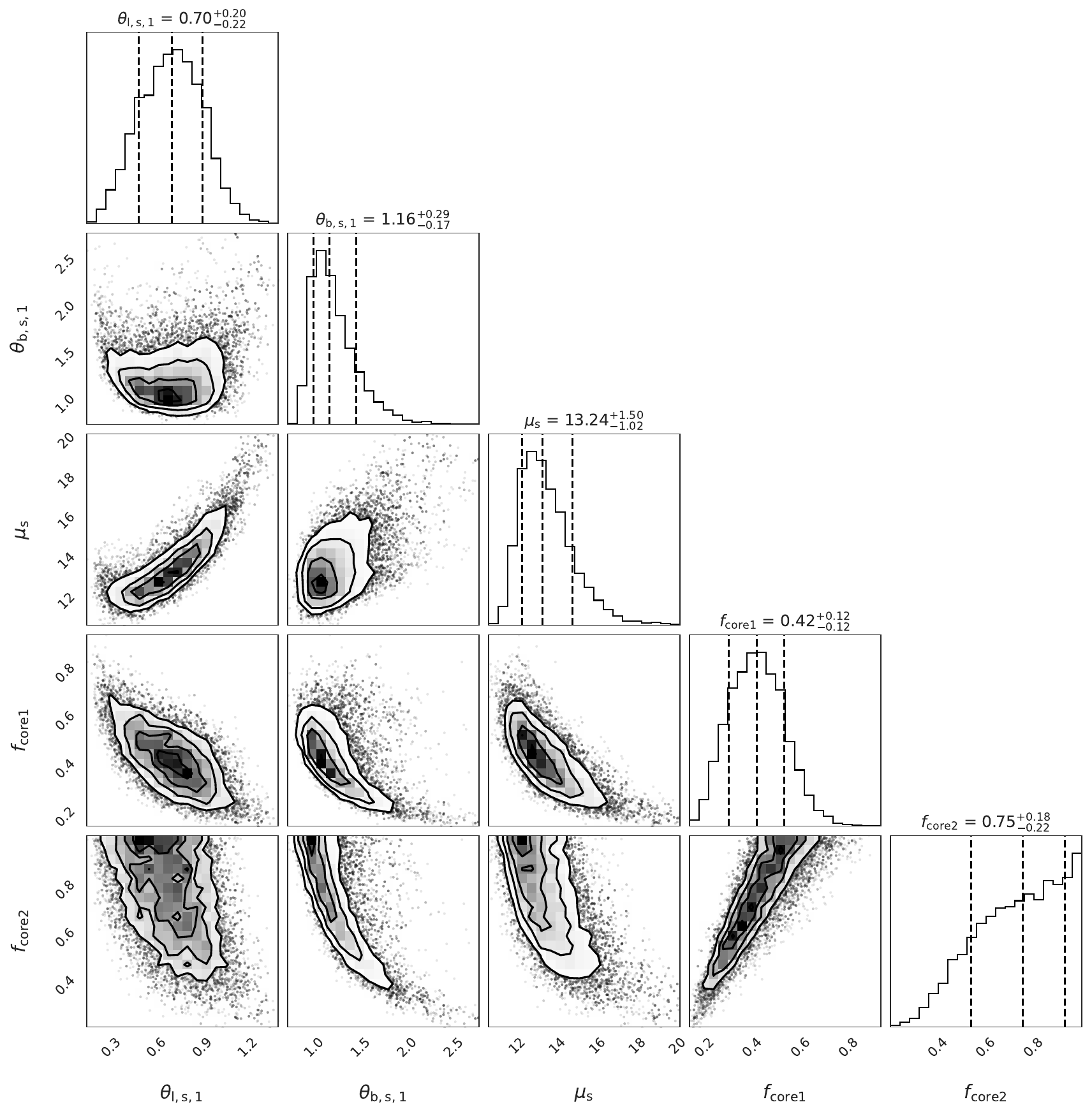}
    \caption{Posterior distribution of the estimated model parameters for the ESE in 2015 using Fiedler's model. Index '1' in parameter name means that this value was estimated for the lowest frequency, i.e. 8.2~GHz. Other indices correspond to higher frequencies in the set in ascending order. Dashed lines represent median and $\pm 1\sigma$.}
    \label{fig:posterior_ese_2015}
\end{figure*}

\begin{figure*}
    \centering
    \includegraphics[width=1\linewidth]{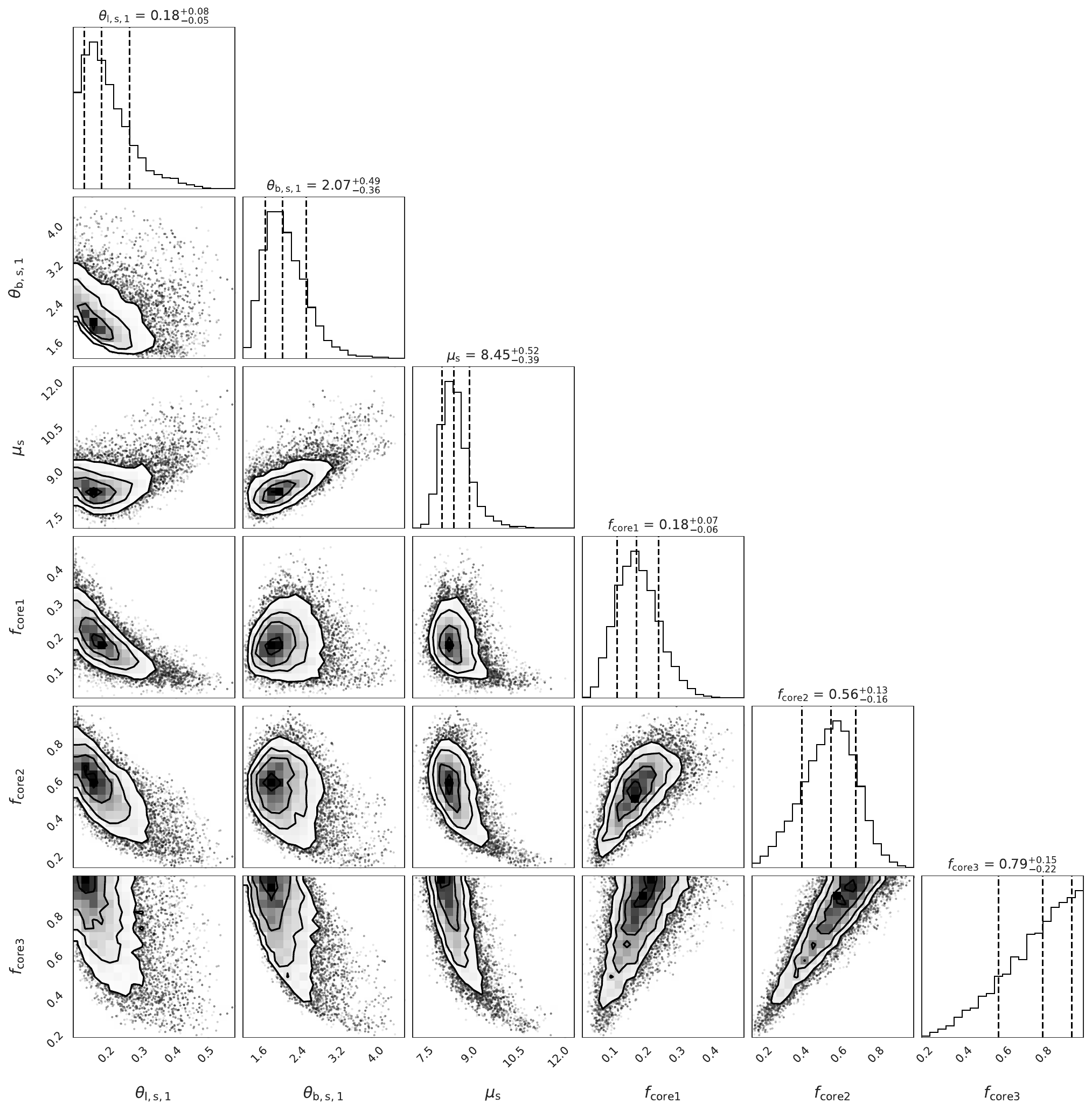}
    \caption{Posterior distribution of the estimated model parameters for the ESE in 2020 using Fiedler's model. Index '1' in parameter name means that this value was estimated for the lowest frequency, i.e. 4.7~GHz. Other indices correspond to higher frequencies in the set in ascending order. Dashed lines represent median and $\pm 1\sigma$.}
    \label{fig:posterior_ese_2020}
\end{figure*}

\begin{figure*}
    \centering
    \includegraphics[width=1\linewidth]{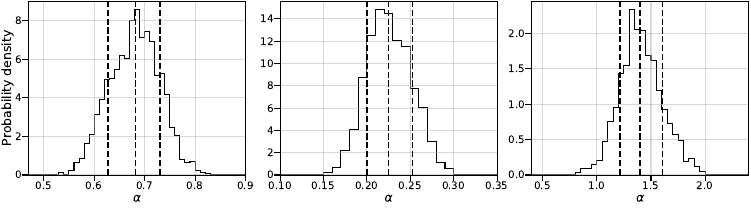}\vspace*{0cm}
    \caption{Posterior distributions of the parameter $\alpha$ estimated using Clegg's model for the ESE in 2011, 2015, and 2020 at 11.2~GHz (left, middle, and right, respectively). Dashed lines represent median and $\pm 1\sigma$.}
    \label{fig:alphas}
\end{figure*}

% Don't change these lines
\bsp % typesetting comment
\label{lastpage}
\end{document}